\pdfoutput=1
\documentclass[11pt]{article}
\usepackage[margin=1in]{geometry}
\usepackage{amssymb}
\usepackage{courier}
\usepackage{times}
\usepackage{caption}
\usepackage{subcaption}
\usepackage{amsfonts}
\usepackage{tikz}
\usepackage{amsmath}
\usepackage{enumitem}
\usetikzlibrary{positioning, shapes, arrows.meta, trees, shapes.geometric, decorations.pathreplacing, calc}
\usepackage{lscape}

\newtheorem{definition}{Definition}

\usepackage[]{graphicx}
\chardef\bslash=`\\

\def\bSig\mathbf{\Sigma}

\renewcommand{\Pr}{\operatorname{P}}

\usepackage{url,amsfonts,threeparttable,adjustbox,longtable,booktabs,multirow,tabularx,dcolumn,bbm,xcolor}

\usepackage[natbibapa]{apacite}
\definecolor{Rred}{HTML}{F8766D}
\definecolor{Rgreen}{HTML}{00BA38}
\definecolor{Rblue}{HTML}{619CFF}

\definecolor{Dogs1}{HTML}{3B9AB2}
\definecolor{Dogs2}{HTML}{6BB1C1}
\definecolor{Dogs3}{HTML}{BDC367}
\definecolor{Dogs4}{HTML}{E6C019}
\definecolor{Dogs5}{HTML}{E49100}
\definecolor{Dogs6}{HTML}{F21A00}

\usepackage{hyperref}
\usepackage{cleveref}
\hypersetup{colorlinks=true, linkcolor=blue, citecolor=blue, urlcolor=blue}

\title{Doubly-robust causal inference for multi-valued treatment regimes with recurrent neural networks}
\author{Jason Poulos\thanks{Division of Endocrinology, Brigham and Women's Hospital, 221 Longwood Avenue, Boston, Massachusetts 02115, USA. Present address: Independent researcher. E-mail: \href{mailto:poulos@berkeley.edu}{poulos@berkeley.edu}}}
\date{}

\begin{document}

\maketitle

\begin{abstract}
We propose a doubly-robust method to estimate the causal effects of multi-valued treatment regimes on patient outcomes. A multi-valued treatment regime assigns personalized treatments from a set of competing options based on a patient’s history. Existing methods for optimal treatment regimes are largely limited to binary or continuous treatments. We extend the longitudinal targeted minimum loss-based estimator (LTMLE) to accommodate multi-valued treatments by modeling treatment assignments with a multinomial model. We further integrate recurrent neural networks (RNNs) into the LTMLE framework as learners for the outcome, treatment, and censoring models, and compare this approach to ensemble machine learning. Simulation studies show that LTMLE with either ensemble machine learning or RNNs is approximately unbiased, with close to nominal coverage, whereas sequential g-computation with ensemble machine learning is substantially biased under a dynamic rule. A multinomial treatment model performs similarly to a series of binomial regressions with ensemble machine learning, but yields more stable inference when the treatment model is an RNN.
\end{abstract}

\noindent\textbf{Keywords:} causal inference, doubly-robust methods, multi-valued treatments, recurrent neural networks, treatment regimes

\section{Introduction}
Evaluating the long-term safety of different treatment regimes on clinically significant outcomes poses a methodological challenge, especially when randomized controlled trials (RCTs) are not feasible. Making reliable causal inferences becomes increasingly complex when studying multiple competing treatment options, moving beyond the binary or continuous treatments typically examined in observational studies. As the number of treatment options increases, the risk of violating the positivity assumption --- a condition ensuring that every patient has a nonzero probability of receiving each possible treatment given their covariates --- also rises. Such violations can limit the ability to make reliable inferences about treatment safety.

A treatment regime is defined as a rule that, at each time point, assigns a specific treatment from a set of possible treatments, as a function of patient characteristics, thus ensuring a ``personalized'' treatment approach \citep{chakraborty2013statistical}. The central question revolves around the methodology for determining the optimal treatment regimes that, on average, result in optimal patient outcomes. However, existing methodologies primarily address scenarios limited to binary or continuous single treatments, overlooking the complexity introduced by scenarios involving multiple competing (multi-valued) treatments \citep{young2011comparative,schomaker2019using,diaz2021nonparametric}, although recent work has begun to address multi-valued time-varying treatments \citep{martin2026nonparametric,susmann2024longitudinal}. In particular, settings where multi-valued treatments are measured at multiple time points are especially prone to violations of the positivity assumption, as the probability of receiving specific treatment sequences can become extremely small or zero for certain covariate patterns \citep{petersen2012diagnosing,mcclean2026comparing}. This challenge is compounded by the curse of dimensionality in longitudinal settings, where the number of possible treatment sequences grows exponentially with the number of time points, making it increasingly difficult to ensure adequate representation of all possible treatment patterns in finite samples \citep{robins2000marginal,van2006targeted}. This sparsity can lead to biased or highly variable estimates of treatment effects, making it challenging to draw valid causal inferences \citep{westreich2010invited}.

These challenges motivate the use of doubly-robust estimators, which provide some protection against model misspecification by requiring that either the treatment model or outcome model is correctly specified to ensure consistent estimation \citep{bang2005doubly,kang2007demystifying,benkeser2017doubly}. Targeted minimum loss-based estimation (TMLE) \citep{van2011targeted,munoz2012population,van2012targeted} is a doubly-robust method that combines initial outcome estimates with functions of estimated treatment and censoring probabilities. Unlike other doubly-robust methods that solve estimating equations, TMLE minimizes a log-likelihood loss function to reduce the bias in the causal parameter estimate. This approach allows for flexible specification of both the outcome and treatment models, including the use of machine learning methods, while maintaining asymptotic efficiency \citep{lee2010improving,austin2012using,pirracchio2015improving}. However, the conventional implementation of TMLE for multi-valued treatments, which relies on multiple binary regressions to estimate treatment probabilities, can compromise efficiency when these models are misspecified. Addressing this limitation, \citet{poulos2024targeted} demonstrated in a cross-sectional intention-to-treat analysis that estimating a single multinomial treatment model, rather than separate binary regressions, leads to efficiency gains in the setting of multi-valued treatments. Their findings highlight the importance of properly modeling the underlying probability distribution of the treatment mechanism, and a recent comparison of estimators for multiple treatment groups similarly favors doubly-robust estimation with machine learning \citep{chen2025comparison}.

Building on these developments, two main frameworks have emerged for analyzing longitudinal data with time-varying treatments: Longitudinal TMLE (LTMLE) and the Longitudinal Modified Treatment Policy (LMTP) framework. LTMLE \citep{stitelman2012general,decker2014semiparametric,schnitzer2014effect,lendle2017ltmle} extends the TMLE methodology to longitudinal settings by iteratively fitting models across time points, and has been successfully applied to comparative effectiveness research \citep{kreif2017comparative}, HIV treatment strategies \citep{schomaker2019using}, and survival analysis \citep{tran2019double,sofrygin2019targeted}. The LMTP framework \citep{diaz2021nonparametric,diaz2022causal,hoffman2023introducing} instead focuses on interventions that depend on the natural value of treatment, offering a more flexible approach for handling multi-valued or continuous treatments \citep{munoz2012population,young2014identification}. While both frameworks share common assumptions of sequential randomization and positivity \citep{robins2000marginal}, they differ in their approach to handling treatment positivity and the specificity of treatment effects they estimate. LMTP's key advantage lies in its ability to mitigate empirical positivity violations by not intervening on observations associated with low-probability natural treatment values \citep{diaz2021nonparametric}, making it particularly useful in observational studies where certain treatment values may be rare or implausible for some covariate combinations \citep{petersen2012diagnosing}. In contrast, LTMLE, while powerful for comparing specific treatment regimes, has traditionally focused on binary treatments and intervention-specific mean differences \citep{petersen2014targeted}. We build on LTMLE because clinical questions often concern fixed or diagnosis-specific protocols, such as the static and dynamic antipsychotic rules in Section~\ref{motivating}, rather than shifts relative to the natural value of treatment; stochastic rules that depend on the previous treatment, as in Section~\ref{simulations}, can also be evaluated within LTMLE.

\subsection{Extending LTMLE with deep learning for multi-valued treatments}

A contribution of this paper is the application of recurrent neural networks (RNNs) to implement LTMLE for multi-valued treatments with a multinomial treatment model. RNNs are a class of neural networks designed to process sequences of data by leveraging their inherent sequential nature \citep{graves2012}. RNNs share model parameters across time periods, allowing them to capture long-term dependencies and adapt to varying sequence lengths --- crucial for longitudinal data where time points may differ across subjects. By maintaining a hidden state that is updated at each step, RNNs can retain information from past inputs, making them adept at tasks like time-series prediction where the order of data points is critical. This approach addresses key limitations of existing methods for estimating treatment effects in complex longitudinal settings. Previous work on LTMLE has primarily focused on binary treatments and relied on traditional statistical models that may struggle with high-dimensional data and complex temporal dependencies \citep{petersen2014targeted,schomaker2019using}. While some recent studies have applied deep learning to causal inference, these have mostly been limited to static settings \citep{bica2020estimating,vowels2021targeted,mi2021deep}, or to single outcomes or treatments in longitudinal settings \citep{poulos2021rnn, li2021g, melnychuk2022causal, xiong2024gtransformer, frauen2025modelagnostic}. Most closely related, \citet{frauen2023estimating} integrate neural networks with iterated g-computation and targeting for time-varying binary treatments, and \citet{shirakawa2024longitudinal} combine a transformer-based initial estimator with LTMLE for dynamic treatment policies; we extend this line of work to multi-valued treatments with a multinomial treatment model, right-censoring, and static, dynamic, and stochastic rules.

Our method utilizes a single multinomial model for multi-valued treatments at each time point, rather than multiple binary regressions, inspired by recent work on cross-sectional TMLE \citep{poulos2024targeted}. This approach properly accounts for the probability structure of mutually exclusive treatments --- enforcing that probabilities sum to one across treatment options and capturing correlations between different treatment choices. Integrating deep learning with LTMLE can offer several advantages in general: improved handling of high-dimensional covariates and non-linear relationships, better modeling of temporal dependencies, and increased model flexibility. Our approach uses an RNN to flexibly model the sequential outcome process, and we analogously model the treatment and censoring mechanisms with an RNN.

In simulation studies, the multinomial treatment model yields more stable inference than separate binomial models when the treatment models are RNNs, and LTMLE with RNNs performs similarly to LTMLE with ensemble machine learning, while g-computation with RNN outcome regressions avoids the bias that ensemble machine learning induces under a dynamic rule. By combining LTMLE's statistical properties with deep learning's modeling power, our approach provides robust treatment effect estimates in complex longitudinal settings, advancing causal inference for time-varying treatments. In Section \ref{estimators}, we present our methodological framework, including the formulation of the estimators and our RNN-based implementation. Section \ref{motivating} provides a motivating example from antipsychotic safety research that illustrates the challenges of analyzing multi-valued treatment regimes in practice. Section \ref{simulations} details our simulation study design (with the data-generating process in Appendix A.1) and presents simulation results comparing our method against traditional approaches across different treatment rules and scenarios. We conclude by discussing implications for causal inference methodology and directions for future research.

\section{Motivating example: safety effects of antipsychotic treatment regimes}\label{motivating}

Antipsychotic medications play a crucial role in the management of serious mental illnesses (SMI), including schizophrenia, bipolar I disorder, and treatment-resistant major depressive disorder (MDD). While over 20 antipsychotic medications are available to clinicians, second-generation antipsychotics (SGAs) are among the most frequently prescribed due to their improved side effect profile compared to first-generation antipsychotics (FGAs). However, some SGAs are associated with an elevated risk of cardiometabolic morbidity, including diabetes \citep{correll2015cardiovascular}. This risk adds to an already elevated baseline: people with schizophrenia have about twice the odds of type 2 diabetes as people without the disorder \citep{dong2024schizophrenia}. The evidence base for the relative diabetes risk of different antipsychotics remains limited: clinical trials have primarily focused on intermediate risk factors like weight gain rather than directly measuring diabetes incidence \citep{pillinger2020comparative}. Observational studies examining diabetes outcomes have often used methods that may not adequately control for differences between treatment groups, potentially leading to biased estimates of the relative diabetes risks associated with different antipsychotic medications \citep[e.g.,][]{gianfrancesco2006influence}.

The cross-sectional study in \citet{poulos2023antipsychotics} applied the novel TMLE methodology with a multinomial treatment model first proposed in \citet{poulos2024targeted} to re-examine the effect of commonly used antipsychotic drugs on the risk of diabetes and mortality in a publicly insured U.S. adult population with SMI. This study compared several SGAs (olanzapine, quetiapine, risperidone, and ziprasidone) and one FGA (haloperidol) to aripiprazole as the reference drug. Previous research suggests that haloperidol, less frequently used than SGAs in the U.S. overall, has a favorable cardiometabolic profile relative to SGAs, while aripiprazole, a more widely used drug, has lower cardiometabolic risk relative to other SGAs \citep[e.g.,][]{huhn2019comparative,pillinger2020comparative}. However, their comparative risk has rarely been directly assessed. The authors found a reduction in diabetes risk associated with initiating haloperidol, olanzapine, or risperidone treatment relative to initiating aripiprazole treatment. This study was among the first to leverage a doubly-robust
machine-learning based method to assess the relative safety of antipsychotic drugs using observational data. \citet{rubinstein2026bounding} subsequently derived bounds on causal effects when outcomes may be missing informatively, for example because of death or loss of insurance coverage; among older Medicare Advantage enrollees with schizophrenia, the lower diabetes risk associated with haloperidol was robust to a substantial share of informative missingness.

These findings on comparative drug safety underscore the methodological and clinical challenges in determining optimal antipsychotic treatment regimes. In particular, the unexpected diabetes risk findings from \citet{poulos2023antipsychotics} and \citet{agniel2024revisiting} suggest that conventional assumptions about antipsychotic safety profiles may need reconsideration. To address this complexity, three distinct approaches to antipsychotic drug assignment have emerged: static, dynamic, and stochastic treatment regimes. Static regimes assign a specific treatment independent of patient characteristics, such as prescribing quetiapine for all patients with bipolar disorder \citep{huhn2019comparative}. While straightforward to implement, this approach may not adequately address individual patient needs. Dynamic regimes adapt treatment based on time-varying patient characteristics \citep{murphy2003optimal,moodie2007demystifying,chakraborty2013statistical}, for example, switching from quetiapine to risperidone if cardiometabolic concerns arise. Stochastic regimes incorporate probabilistic decision-making, where treatment modifications follow predetermined likelihoods \citep{kim2021incremental,mcclean2026comparing}, offering flexibility while maintaining protocol consistency. The choice among these regime types depends on factors including SMI diagnosis, individual cardiometabolic risk factors, intended treatment duration, and symptom control requirements \citep{stroup2011randomized}.

\section{Doubly-robust estimation}\label{estimators}

Doubly-robust estimation methods combine outcome regression and propensity score approaches to provide consistent estimates of causal effects even if one of the component models is misspecified. This section describes two singly-robust estimators for longitudinal data, sequential g-computation and inverse probability of treatment weighting (IPTW), the doubly-robust LTMLE, and incorporating sequential deep learning models like RNNs.

\subsection{Notation}
Consider a cohort of $n$ subjects observed at baseline ($t=0$) and at subsequent discrete follow-up times $t = 1, \ldots, T$, where $T$ delineates the end of the observation period. It is possible for some subjects to be censored and not followed up to time $T$. The dataset at each time point $t$ includes the outcome variable $Y_t$, an intervention variable $A_t$, a set of $q$ time-dependent covariates denoted by $\mathbf{L}_t = \{L_t^1, \ldots, L_t^q\}$, and a censoring indicator $C_t$. Baseline variables, measured at $t = 0$, are represented as $\mathbf{L}_0 = \{L_0^1, \ldots, L_0^q\}$. Treatment and covariate histories for subject $i$ up to and including time $t$ are denoted as $\bar{\mathbf{A}}_{t,i} = (A_{0,i}, \ldots, A_{t,i})$ and $\bar{\mathbf{L}}_{t,i} = ({L_{0,i}^1, \ldots, L_{0,i}^q}, \ldots, {L_{t,i}^1, \ldots, L_{t,i}^q})$, respectively. The censoring indicator $C_t$ is 0 if a subject is censored in the interval $(t-1, t]$, and 1 otherwise, signifying that the individual remains uncensored until time $t$. Within each time point the variables are ordered $\mathbf{L}_t, A_t, C_t, Y_t$, with $C_0 \equiv 1$, and $\mathbf{L}_0$ includes the time-invariant baseline characteristics. The outcome $Y_t \in \{0,1\}$ indicates an event by time $t$ and is absorbing ($Y_s=1$ for $s>t$ once $Y_t=1$; $Y_0 \equiv 0$), as is censoring. 

The counterfactual outcome $Y_{t,i}^{\bar{\mathbf{a}}_t}$ depicts the potential outcome at time $t$ for subject $i$ given the treatment history $\bar{\mathbf{A}}_{t,i} = \bar{\mathbf{a}}_t$, which may differ from the actual treatment received. The notation $\bar{\mathbf{A}}_t = \bar{\mathbf{d}}_t$ refers to the treatment history up to and including time $t$. Under a static treatment rule $\bar{\mathbf{d}}_t^1$, the assignment depends only on the baseline diagnosis (not on time-varying covariates) and remains constant over time, with counterfactual covariates denoted as $\mathbf{L}_{t,i}^{\bar{\mathbf{d}}_t^1}$. For a dynamic treatment regime $\bar{\mathbf{d}}_t^2$, the counterfactual covariates $\mathbf{L}_{t,i}^{\bar{\mathbf{d}}_t^2}$ evolve in response to prior treatments $\bar{\mathbf{A}}_{t-1,i}=\bar{\mathbf{d}}_{t-1}^2$, each assigned according to the covariate history at that time, and $A_{t,i}$ is then assigned according to $\bar{\mathbf{L}}_{t,i}$. Lastly, a stochastic rule $\bar{\mathbf{d}}_t^3$ introduces a probabilistic element, where treatment at each time point $t > 0$ is assigned probabilistically conditional on the previous treatment, creating a potential for random shifts in treatment assignment according to a Multinomial distribution. The counterfactual covariates under the stochastic rule are denoted as $\mathbf{L}_{t,i}^{\bar{\mathbf{d}}_t^3}$.

The target quantity of interest, denoted as $\psi_t$, is the expected value of the outcome $Y$ at time point $t$ under an intervention rule $\bar{\mathbf{d}}_t$, which assigns treatment $A_t$ (deterministically for the static and dynamic rules, randomly for the stochastic rule) as a function of the observed history, under no censoring.

\subsection{The g-formula}
The quantity of interest is the expected value of an outcome at each time point, which is identifiable through the g-formula \citep{van2018targetedgcomp}. Formally, the target quantity \(\psi_t\) for any time point \( t \) (with \( t \leq T \)) is given by:
\begin{align}\label{g-formula}
    \psi_t = \operatorname{E}(Y_t^{\bar{\mathbf{d}}_t}) &= \int_{\mathcal{L}_t} \bigg[ \operatorname{E}\{Y_t \mid \bar{\mathbf{A}}_{t} = \bar{\mathbf{d}}_{t}, \bar{\mathbf{C}}_{t} = 1, \bar{\mathbf{L}}_t = \bar{\mathbf{l}}_t\} \notag \\
    &\phantom{=} \times \prod_{k=0}^{t} f(\mathbf{l}_{k} \mid \bar{\mathbf{A}}_{k-1} = \bar{\mathbf{d}}_{k-1}, \bar{\mathbf{C}}_{k-1} = 1, \bar{\mathbf{L}}_{k-1} = \bar{\mathbf{l}}_{k-1}) \bigg] d\bar{\mathbf{l}}_t,
\end{align}
where $\mathcal{L}_t$ is the support of $\bar{\mathbf{L}}_t$ and $k$ iterates over all time points leading up to $t$. The function $f(\cdot)$ represents the conditional density of the time-varying covariates $\mathbf{L}_t$; for $k=0$ the conditioning set is empty, so $f(\mathbf{l}_0)$ is the marginal density of the baseline covariates. For the stochastic rule, the indicator $\bar{\mathbf{A}}_t=\bar{\mathbf{d}}_t$ is replaced by integration over the rule's assignment probabilities. The variables $\mathbf{L}_t$ can be ordered as $\{L_t^1, \ldots, L_t^q\}$, allowing for the joint distribution to be expressed sequentially according to this order.

The validity of \eqref{g-formula} relies on several assumptions. The first assumption is \textit{consistency}, expressed as $Y_{t}^{\bar{\mathbf{d}}_{t}} = Y_t$ for any instance where $\bar{\mathbf{A}}_{t} = \bar{\mathbf{d}}_{t}$, $\bar{\mathbf{C}}_{t} = 1$, and $\bar{\mathbf{L}}_{t} = \bar{\mathbf{l}}_{t}$, which posits that the counterfactual outcome of a patient given their factual treatment history should coincide with the observed outcome. The second assumption, termed \textit{(sequential) unconfoundedness}, posits that the counterfactual outcome under the specified treatment regime is statistically independent of the treatment assignment observed, conditional on being uncensored and the observed covariate history, and is represented by $Y_{t}^{\bar{\mathbf{d}}_t} \perp\!\!\!\perp A_k \mid \bar{\mathbf{L}}_{k}, \bar{\mathbf{A}}_{k-1}=\bar{\mathbf{d}}_{k-1}, \bar{\mathbf{C}}_{k-1}=1$ and $Y_{t}^{\bar{\mathbf{d}}_t} \perp\!\!\!\perp C_k \mid \bar{\mathbf{L}}_{k}, \bar{\mathbf{A}}_{k}=\bar{\mathbf{d}}_{k}, \bar{\mathbf{C}}_{k-1}=1$ for each $k \le t \le T$ and all possible treatment regimes $\bar{\mathbf{d}}$ and covariate histories $\bar{\mathbf{l}}_{t}$. The final assumption, known as \textit{positivity}, asserts that there exists a non-zero probability for each individual to receive the treatment according to the regime $\bar{\mathbf{d}}_t$, based on their history of covariates up to that juncture; formally, this is stated as $\Pr(A_{t} = d_{t}, C_t = 1 \mid \bar{\mathbf{L}}_{t} = \bar{\mathbf{l}}_{t}, \bar{\mathbf{A}}_{t-1} = \bar{\mathbf{d}}_{t-1}, \bar{\mathbf{C}}_{t-1}=1) > 0$ for every $k \le t$ (with $t$ replaced by $k$) and all $\bar{\mathbf{d}}_{t}, \bar{\mathbf{l}}_{t}$ in the context where $\Pr(\bar{\mathbf{L}}_{t} = \bar{\mathbf{l}}_{t}, \bar{\mathbf{A}}_{t-1} = \bar{\mathbf{d}}_{t-1}, \bar{\mathbf{C}}_{t-1}=1) > 0$.

\subsection{Sequential g-computation} 
The sequential g-formula, a variant of \eqref{g-formula}, iteratively marginalizes the distribution with respect to time-dependent covariates \(\mathbf{L}_t\), given the assigned treatment rule $\bar{\mathbf{d}}_t$. The estimator's formula at any time point \(t\) is:
\begin{equation}\label{sequential-g}
\operatorname{E} \left( Y_{t}^{\bar{\mathbf{d}}_t} \right) = \operatorname{E}_{\mathbf{L}_0} \circ \operatorname{E}_{\mathbf{L}_1} \circ \cdots \circ \operatorname{E}_{\mathbf{L}_{t}} \left[ Y_t \mid \bar{\mathbf{A}}_t = \bar{\mathbf{d}}_t, \bar{\mathbf{C}}_t = 1, \bar{\mathbf{L}}_t \right],
\end{equation}
where each inner expectation conditions on the history up to that time with treatments set by the rule and no censoring, and \(\circ\) denotes function composition. The expectations are evaluated sequentially backward from time \(t\) to 0. This notation is equivalent to the nested conditional expectations evaluated at each time point, with each expectation conditioned on the treatment and covariate history available at that time.

The process to estimate the target parameter using \eqref{sequential-g} includes:
\begin{enumerate}
    \item For $k = t, t-1, \ldots, 0$, regress $\hat{Y}_{k+1}^{\bar{\mathbf{d}}_t}$ on $(\bar{\mathbf{A}}_{k}, \bar{\mathbf{L}}_k)$ among subjects uncensored through $k$ and event-free through $k-1$ (subjects with an event by $k$ have pseudo-outcome 1), where $\hat{Y}_{t+1}^{\bar{\mathbf{d}}_t} \equiv Y_t$ is the observed outcome and, for $k<t$, $\hat{Y}_{k+1}^{\bar{\mathbf{d}}_t}$ is the prediction from the preceding iteration $k+1$.
    \item Set $A_{k}=d_{k}$ based on the assigned rule (earlier treatments keep their observed values) and use the model from the first step to obtain the prediction $\hat{Y}_k^{\bar{\mathbf{d}}_t}$. Under the stochastic rule, which assigns $A_k$ with known probabilities $g^*_k(a \mid A_{k-1})$ for $k \ge 1$, the prediction is instead averaged over treatments, $\hat{Y}_k^{\bar{\mathbf{d}}_t} = \sum_{a=1}^{J} g^*_k(a \mid A_{k-1})\,\hat{\operatorname{E}}[\hat{Y}_{k+1}^{\bar{\mathbf{d}}_t} \mid A_k = a, \bar{\mathbf{A}}_{k-1}, \bar{\mathbf{L}}_k, \bar{\mathbf{C}}_k = 1]$, and $A_0$ is left at its observed value.
    \item Compute $\hat{\psi}_t$ by averaging $\hat{Y}_0^{\bar{\mathbf{d}}_t}$ across all $n$ individuals. For g-computation we report Wald intervals based on \eqref{eq:IC} evaluated at the untargeted fits. In \eqref{eq:IC}, $H_s$ is the inverse of the estimated probability of following the rule and remaining uncensored through time $s$ (the clever covariate of the LTMLE, defined in Section~\ref{ltmle}); for g-computation it is formed from the multinomial treatment and censoring models of the same learner (Super Learner or RNN); because the plug-in estimator with data-adaptive learners is not asymptotically linear, these intervals are approximate and expected to undercover.
\end{enumerate}

\subsection{Inverse Probability of Treatment Weighting (IPTW)}
IPTW employs weights to create a pseudo-population where the treatment assignment across time is independent of previously measured covariates \citep{robins2000marginal,young2014identification}. This method helps to estimate the causal effect of a treatment by creating a weighted dataset that mimics a randomized controlled trial. The IPTW estimator is expressed as:
\begin{align*}
\hat{\psi}_t &= \frac{\sum_{i=1}^{n} \hat{w}_{t,i} Y_{t,i}}{\sum_{i=1}^{n} \hat{w}_{t,i}}, \qquad \hat{w}_{t,i} = \left\{\frac{\mathbbm{1}(\bar{\mathbf{A}}_{t} = \bar{\mathbf{d}}_t, \bar{\mathbf{C}}_{t} = 1)}{\hat{g}_t}\right\}_i, \\
\hat{g}_t &= \prod_{k=0}^{t} \hat{\Pr}(A_{k} = d_k \mid \bar{\mathbf{L}}_{k}, \bar{\mathbf{A}}_{k-1} = \bar{\mathbf{d}}_{k-1}, \bar{\mathbf{C}}_{k-1} = 1) \\
&\phantom{=} \times \prod_{k=1}^{t} \hat{\Pr}(C_k = 1 \mid \bar{\mathbf{L}}_{k}, \bar{\mathbf{A}}_{k} = \bar{\mathbf{d}}_{k}, \bar{\mathbf{C}}_{k-1} = 1).
\end{align*}
In this expression, \( \bar{\mathbf{A}}_t \) encapsulates the sequence of treatments administered up to and including time point \( t \), and \( \bar{\mathbf{d}}_t \) represents the predetermined intervention rule. The indicator function \( \mathbbm{1}(\bar{\mathbf{A}}_t = \bar{\mathbf{d}}_t, \bar{\mathbf{C}}_{t} = 1 )\) yields 1 when the treatment history \( \bar{\mathbf{A}}_t \) aligns with the intervention rule \( \bar{\mathbf{d}}_t \) and the subject is uncensored up to time $t$, and it is 0 otherwise. The denominator $\hat{g}_t$ quantifies the joint probability of being administered the treatments $\bar{\mathbf{d}}_t$ in accordance with the intervention rule and remaining uncensored, conditional on the historical covariate trajectory \( \bar{\mathbf{L}}_{k} \) and previous treatments \( \bar{\mathbf{A}}_{k-1} \). Note that treatment is first assigned and then the censoring indicator is recorded. Consequently, the probability of remaining uncensored is conditioned on both the covariate history \(\bar{\mathbf{L}}_k\) and the realized treatment at time \(k\) (i.e., \(\bar{\mathbf{A}}_k = \bar{\mathbf{d}}_k\)). This normalized form divides by the sum of the weights instead of $n$. For the stochastic rule, the indicator $\mathbbm{1}(A_k = d_k)$ is replaced by $g^*_k(A_k \mid A_{k-1})$ for $k \ge 1$ (and by 1 at $k=0$) and $\hat{\Pr}(A_k = d_k \mid \cdot)$ by $\hat{\Pr}(A_k \mid \cdot)$, so that the weight is a cumulative density ratio; deterministic rules are the special case $g^*_k(a \mid \cdot) = \mathbbm{1}(a = d_k)$.

\subsection{Longitudinal targeted minimum loss-based estimator (LTMLE)}\label{ltmle}
The LTMLE framework involves iteratively fitting models for outcome, censoring, and treatment at each time point. The models are fit on all uncensored subjects and evaluated on those who follow a treatment rule. Like the TMLE, the estimator remains consistent if either the outcome models or the treatment/censoring models are estimated consistently. The steps for estimating the expected value of the outcome $Y$ at each time point $t$ under an intervention rule $\bar{\mathbf{d}}_t$ are updated as follows:

\begin{enumerate}
    \item Regress $Y_t$ on the observed history $(\bar{\mathbf{A}}_{t}, \bar{\mathbf{L}}_t)$ among subjects who are uncensored through $t$ and event-free through $t-1$. This model specifies the conditional expectation of the outcome $Y_t$ considering the history of treatment and covariates; as in sequential g-computation, the regression is repeated backward for $k=t,\ldots,0$, using the targeted prediction from step $k+1$ (Step 3) as the outcome.

    \item Predict the outcome at time $t$ using the estimated model, plugging in $A_{t} = d_{t}$ to obtain $\hat{Y}_t^{\bar{\mathbf{d}}_t}$; at each later step $k$ of the recursion, $\hat{Y}_k^{\bar{\mathbf{d}}_t}$ is obtained in the same way with $A_k = d_k$. The stochastic rule does not assign a single treatment but draws $A_k$ from the known probabilities $g^*_k(a \mid A_{k-1})$, so there is no single value to plug in. Instead, the model is evaluated at each of the $J$ treatment values $A_k = a$, and the $J$ predictions are averaged with weights $g^*_k(a \mid A_{k-1})$, $\hat{Y}_k^{\bar{\mathbf{d}}_t} = \sum_{a=1}^{J} g^*_k(a \mid A_{k-1})\,\hat{\operatorname{E}}[\hat{Y}_{k+1}^{\bar{\mathbf{d}}_t} \mid A_k = a, \bar{\mathbf{A}}_{k-1}, \bar{\mathbf{L}}_k, \bar{\mathbf{C}}_k = 1]$, as in Step 2 of sequential g-computation; $A_0$ keeps its observed value.

    \item Update the initial estimates of the target parameter $\psi_t$ by incorporating a \emph{clever covariate} $H_t(\bar{\mathbf{A}}, \bar{\mathbf{C}}, \bar{\mathbf{L}})$, computed as:
    \begin{equation*}
    H_t = \prod_{s=0}^{t} \frac{\mathbbm{1}(A_s = d_s, C_s = 1)}{\hat{\Pr}(A_s = d_s \mid \bar{\mathbf{L}}_s, \bar{\mathbf{A}}_{s-1}=\bar{\mathbf{d}}_{s-1}, \bar{\mathbf{C}}_{s-1}=1) \cdot \hat{\Pr}(C_s = 1 \mid \bar{\mathbf{A}}_s = \bar{\mathbf{d}}_s, \bar{\mathbf{L}}_s, \bar{\mathbf{C}}_{s-1}=1)}
    \end{equation*}
    This function represents the inverse probability weights for each individual, where each weight is the inverse of the probability of following the treatment regime $\bar{\mathbf{d}}_t$ up to time $t$, conditional on the history of covariates $\bar{\mathbf{L}}_t$ and the absence of censoring, $\bar{\mathbf{C}}_t = \prod_{s=0}^t C_s$, with $\Pr(C_0=1)\equiv 1$ since there is no censoring at baseline. Under the stochastic rule, the numerator indicator for treatment is replaced by $g^*_s(A_s \mid A_{s-1})$ (1 at $s=0$) and the treatment probability in the denominator is evaluated at the observed $A_s$, as for IPTW. At each step $k = t,\ldots,0$ of the backward recursion, $\hat{\varepsilon}_k$ is obtained from an intercept-only logistic regression of $\hat{Y}^{*\bar{\mathbf{d}}_t}_{k+1}$ ($\hat{Y}^{*\bar{\mathbf{d}}_t}_{t+1}\equiv Y_t$) with offset equal to the logit of the step-$k$ prediction at the observed history and weights $H_k$, among subjects uncensored through $k$ and event-free through $k-1$, so that the score equation $\sum_i H_{k,i}(\hat{Y}^{*\bar{\mathbf{d}}_t}_{k+1,i} - \hat{Y}^{*\bar{\mathbf{d}}_t}_{k,i}) = 0$ holds. The counterfactual predictions of all at-risk subjects are then updated as $\operatorname{logit}\hat{Y}^{*\bar{\mathbf{d}}_t}_k = \operatorname{logit}\hat{Y}^{\bar{\mathbf{d}}_t}_k + \hat{\varepsilon}_k$ (under the stochastic rule, for each treatment value before averaging over $g^*_k$), and $\hat{Y}^{*\bar{\mathbf{d}}_t}_k$ is used as the outcome at step $k-1$.

    \item Obtain the estimate $\hat{\psi}_t$ by averaging the targeted predictions $\hat{Y}_0^{*\bar{\mathbf{d}}_t}$ over all $n$ individuals.

    \item Compute confidence intervals for $\psi_t$ using the estimated influence curve $\widehat{IC}(\psi_t)$:
    \begin{equation}\label{eq:IC}
    \widehat{IC}(\psi_t) \;=\; \sum_{s=0}^{t} H_s(\bar{\mathbf{A}},\bar{\mathbf{C}},\bar{\mathbf{L}})\,\Big[\hat{Y}_{s+1}^{*\,\bar{\mathbf{d}}_t} - \hat{Y}_{s}^{*\,\bar{\mathbf{d}}_t}\Big] \;+\; \hat{Y}_0^{*\,\bar{\mathbf{d}}_t} \;-\; \hat{\psi}_t~.
    \end{equation}
    Here $H_s(\cdot)$ denotes the product of weights up to time $s$ (analogous to $H_t$ but truncated at $s$), $\hat{Y}_s^{*\bar{\mathbf{d}}_t}$ inside the sum is the targeted step-$s$ regression evaluated at the subject's observed history through $A_s$ (which equals the rule-based prediction whenever $H_s \neq 0$ under a deterministic rule), $\hat{Y}_{t+1}^{*\bar{\mathbf{d}}_t}\equiv Y_t$ is the observed outcome, and the final term $\hat{Y}_0^{*\bar{\mathbf{d}}_t}$ is the targeted baseline prediction under the rule.

    Since subjects are independent, we estimate the asymptotic variance by the empirical variance of the estimated influence curve,
    \begin{equation*}
    \widehat{\operatorname{Var}}\{\widehat{IC}(\psi_t)\} = \frac{1}{n}\sum_{i=1}^{n} \widehat{IC}_i(\psi_t)^2,
    \end{equation*}
    where $i$ indexes subjects; temporal dependence within a subject is captured by the sum over $s$ in \eqref{eq:IC}.
\end{enumerate}

In practice, the initial outcome regression (via either Super Learner or RNN, described next) and the targeted update are interleaved at each backward step $k = t, \ldots, 0$, since the targeted prediction from step $k+1$ is the pseudo-outcome at step $k$; this solves the efficient influence curve estimating equation for $\psi_t$. The nuisance models are fit and evaluated on the same subjects, without cross-fitting. The asymptotically normal $95\%$ confidence interval for $\psi_t$ is given by
\begin{equation*}
\hat{\psi}_t \pm 1.96 \sqrt{\frac{\widehat{\operatorname{Var}}\{\widehat{IC}(\psi_t)\}}{n}}.
\end{equation*}

\subsection{LTMLE with RNNs}
RNNs operate with a hidden state vector that updates with each input in a sequence. This vector stores prior input information, providing a foundation for the network to predict future outcomes. One of the key advantages of RNNs is their parameter sharing across time steps, which not only reduces the number of parameters required but also enables the network to generalize across different time periods. This property is crucial in LTMLE as it allows for the modeling of units over multiple time periods, considering the entire trajectory of each unit's outcome and treatment/censoring probabilities. 

For each time step \( t \), the RNN accepts an input \( \mathbf{x}_t \) and applies a transformation that involves the prior hidden state \( \mathbf{h}_{t-1} \) to compute the current hidden state \( \mathbf{h}_t \). The hidden state encapsulates the information of all prior inputs up to the current time step. Subsequently, \( \mathbf{h}_t \) is utilized to generate a prediction for the output \( \hat{y}_t \). 

Formally, a basic recurrent cell, which we use to illustrate the idea before the gated long short-term memory (LSTM) cells of our implementation, involves three principal equations:
\begin{align}
    \mathbf{a}_t &= \mathbf{b} + \mathbf{W}\mathbf{h}_{t-1} + \mathbf{U}\mathbf{x}_t, \label{eq:a} \\
    \mathbf{h}_t &= f(\mathbf{a}_t), \label{eq:h} \\
    \hat{y}_t &= \sigma(\mathbf{c} + \mathbf{V}\mathbf{h}_t), \label{eq:y} 
\end{align}
where \( \mathbf{b} \) and \( \mathbf{c} \) are bias vectors. The parameters used for computing the hidden state, represented by matrices \( \mathbf{W} \) and \( \mathbf{U} \), and the output prediction, represented by matrix \( \mathbf{V} \), are shared across all time steps, ensuring parameter efficiency and the ability of the network to generalize across different temporal sequences. In \eqref{eq:a}, \( \mathbf{a}_t \) denotes the pre-activation vector at time \( t \), computed by integrating the previous hidden state \( \mathbf{h}_{t-1} \) and the current input \( \mathbf{x}_t \) with the respective weights and bias. Equation \eqref{eq:h} applies the nonlinear activation function \( f \), commonly chosen to be the hyperbolic tangent (tanh), to \( \mathbf{a}_t \) to update the hidden state \( \mathbf{h}_t \), which encapsulates the learned information from the sequence thus far. Equation \eqref{eq:y} uses the updated hidden state \( \mathbf{h}_t \) to predict the outcome \( \hat{y}_t \) at time $t$ by applying another set of weights \( \mathbf{V} \) and bias \( \mathbf{c} \), followed by the output link \( \sigma \), the logistic (binary models) or softmax (categorical treatment model) function.

For each time step \( t \), gradients of the loss function with respect to the parameters in \eqref{eq:a}--\eqref{eq:y} are computed using the back-propagation through time algorithm \citep{goodfellow2016deepRNN}, which are then used to update the network parameters via gradient descent. This iterative training process ensures that the information from the historical data is carried over to the next time step, enabling the RNN to make informed predictions based on the cumulative data. It is this recursive property that allows RNNs to effectively model temporal dependencies and has made them a popular choice for applications like time series analysis and sequence prediction \citep{cinar2017position,zhu2017deep}. The hidden state is initialized at \( \mathbf{h}_0=\mathbf{0} \), and the input and recurrent weight matrices are initialized with Glorot-uniform \citep{glorot2010} and orthogonal initializers, respectively, setting the stage for the network's iterative learning process. LSTMs are a type of RNN architecture designed to mitigate vanishing gradients, which arise from repeated multiplication by the recurrent weights over many time steps and prevent standard RNNs from learning long-term dependencies \citep{schmidhuber1997long, pascanu2013difficulty}.

In our implementation, RNNs serve as the learners within the LTMLE of Section~\ref{estimators}: the outcome regression at each step $k=t,\ldots,0$ of the backward recursion is an LSTM that maps the subject's last 12 months of history ending at $k$ to $\hat{Y}_k^{\bar{\mathbf{d}}_t}$, with only the month-$k$ treatment $A_k$ set to $d_k$ for counterfactual predictions, and each step is initialized from the network of step $k+1$. The baseline characteristics are part of every month's input, so they remain available when month 0 falls outside the window. The 12-month window imposes a finite-memory assumption on the outcome regressions, which holds in our simulation because the DGP is first-order Markov (Appendix A.1); the Super Learner (hereafter, ``SL'') outcome regressions likewise use the baseline characteristics and the current and previous month's covariates and treatments. The treatment and censoring mechanisms are estimated by many-to-many LSTMs that output $\hat{\Pr}(A_t \mid \bar{\mathbf{L}}_t, \bar{\mathbf{A}}_{t-1}, \bar{\mathbf{C}}_{t-1}=1)$ and $\hat{\Pr}(C_t=1\mid\cdot)$ at every month from the history up to that month.

\begin{figure}[htbp]
\centering
\resizebox{\textwidth}{!}{%
\begin{tikzpicture}[
    node distance=1.5cm,
    box/.style={rectangle, draw, minimum width=2cm, minimum height=1cm},
    lstm/.style={rectangle, draw, fill=blue!10, minimum width=1.5cm, minimum height=1.2cm},
    attention/.style={rectangle, draw, fill=orange!10, minimum width=1.5cm, minimum height=1.2cm},
    proc/.style={rectangle, draw, fill=green!10, minimum width=2.5cm},
    data/.style={rectangle, draw, dashed, minimum width=1.5cm},
    arrow/.style={->,>=stealth, thick},
    equation/.style={align=left}
]

\begin{scope}[shift={(-16,6.2)}]
    \node[rotate=90] at (-4,0) {\textbf{\shortstack{Network}}};
    \node[box, align=center, minimum height=1.2cm] (xin) at (-1,0) {\shortstack{Inputs $\mathbf{x}_k$\\(own history)}};
    \node[lstm, align=center, right=0.6cm of xin] (l1) {\shortstack{LSTM (32)\\+ LayerNorm}};
    \node[lstm, dashed, align=center, right=0.6cm of l1] (l2) {\shortstack{LSTM (32)\\+ LayerNorm}};
    \node[attention, dashed, align=center, right=0.6cm of l2] (att) {\shortstack{Causal self-\\attention}};
    \node[proc, align=center, minimum height=1.2cm, right=0.6cm of att] (dense) {\shortstack{Dense (16, ReLU)\\$\oplus\ \mathbf{x}_k$}};
    \node[box, align=center, minimum height=1.2cm, right=0.6cm of dense] (out) {\shortstack{$\sigma$ / softmax\\$\rightarrow \hat{y}$}};
    \draw[arrow] (xin) -- (l1);
    \draw[arrow] (l1) -- (l2);
    \draw[arrow] (l2) -- (att);
    \draw[arrow] (att) -- (dense);
    \draw[arrow] (dense) -- (out);
    \draw[arrow] (l1.north) to[out=45, in=135] node[above] {\small residual} (att.north west);
    \draw[arrow] (xin.south) to[out=-20, in=-160] node[below, pos=0.5] {\small input skip} (dense.south);
    \node[align=left, anchor=north west, text width=17cm] at (-2.3,-2.5) {\small Dashed blocks: treatment and censoring networks only (an output at every month $t=0,\ldots,36$); the outcome networks use one LSTM layer over a 12-month window and its last hidden state.};
\end{scope}

\begin{scope}[shift={(-16,0)}]
    \node[rotate=90] at (-4,0) {\textbf{\shortstack{Feature\\ Processing}}};
    
    \node[proc] (norm) at (-1,0) {
        \begin{tabular}{c}
        Normalization \\
        $\mathbf{V}^3$: $x' = \frac{x - \text{med}(x)}{\text{IQR}(x)}$ \\
        $\mathbf{L}^1$: $x' = \log(1+x)$
        \end{tabular}
    };
    
    \node[proc, right=2cm of norm] (temp) {
        \begin{tabular}{c}
        Temporal Encoding \\
        $t' = \frac{t}{T}$
        \end{tabular}
    };
    
    \draw[arrow] (norm) -- (temp);
    
    \begin{scope}[yshift=-3cm]
        \node[rotate=90] at (-4,1) {\textbf{\shortstack{Outcome\\ Windows \\ (per subject)}}};
        
        \draw[->, >=latex] (-1,-0.5) -- (11.5,-0.5) node[below] {time};
        \foreach \x in {0,1,...,15}
            \draw (\x*0.7 - 0.5,-0.4) -- (\x*0.7 - 0.5,-0.6) node[below=0.1cm] {$t_{\x}$};
        
        \draw[blue!30, fill=blue!10] (-0.5 - 0.35,1) rectangle (-0.5 + 11*0.7 + 0.35,1.5);
        \draw[blue!30, fill=blue!10] (-0.5 + 0.7 - 0.35,0.5) rectangle (-0.5 + 12*0.7 + 0.35,1);
        \draw[blue!30, fill=blue!10] (-0.5 + 1.4 - 0.35,0) rectangle (-0.5 + 13*0.7 + 0.35,0.5);
        
        \node[left] at (-1,1.25) {Ending at $t_{11}$};
        \node[left] at (-1,0.75) {Ending at $t_{12}$};
        \node[left] at (-1,0.25) {Ending at $t_{13}$};
        
        \node[right] at (9,1.5) {Size: 12 months};
        \node[right] at (9,1) {Stride: 1 month};
        \node[right] at (9,0.5) {$k<11$: left-padded, masked};
    \end{scope}
\end{scope}

\begin{scope}[shift={(-16,-6)}]
    \node[lstm, align=center] (outcome) at (-2,0) {\shortstack{Outcome LSTM (one per\\step $k$)}};
    \node[lstm] (censor) at (-2,-1.5) {Censoring LSTM};
    \node[lstm] (treat) at (-2,-3) {Treatment LSTM};
    
    \node[box] (yhat) at (3,0) {$\hat{Y}_k^{\bar{\mathbf{d}}_t}$};
    \node[box] (chat) at (3,-1.5) {$\hat{\Pr}(C_t = 1 \mid \cdot)$};
    \node[box] (ghat) at (3,-3) {$\hat{\Pr}(A_t \mid \bar{\mathbf{L}}_t, \bar{\mathbf{A}}_{t-1})$};
    
    \node[box] (clever) at (6,-1.5) {Clever Covariate $H_t$};
    \node[box] (update) at (9,-1.5) {TMLE Update};
    
    \draw[arrow] (outcome) -- (yhat);
    \draw[arrow] (censor) -- (chat);
    \draw[arrow] (treat) -- (ghat);
    \draw[arrow] (yhat) -- (clever);
    \draw[arrow] (chat) -- (clever);
    \draw[arrow] (ghat) -- (clever);
    \draw[arrow] (clever) -- (update);
    
    \node[below=0.5cm of update] {
        $\hat{\psi}_t = \frac{1}{n}\sum_{i} \hat{Y}^{*\bar{\mathbf{d}}_t}_{0,i}$
    };
\end{scope}

\node[above=1.3cm of l2] {\textbf{(a) LSTM Architectures}};
\node[above=0.2cm of norm] {\textbf{(b) Data Preprocessing Pipeline}};
\node[above=0.2cm of outcome] {\textbf{(c) LTMLE Integration with LSTMs}};

\end{tikzpicture}%
}
\caption{Integration of RNNs into the LTMLE framework, showing (a) the LSTM architectures (treatment/censoring: two LSTM layers with causally masked self-attention; outcome: one LSTM layer over a 12-month window), (b) the data preprocessing pipeline with feature scaling and per-subject outcome windows, and (c) the integration with TMLE for computing targeted updates.}
\label{fig:rnn_ltmle}
\end{figure}
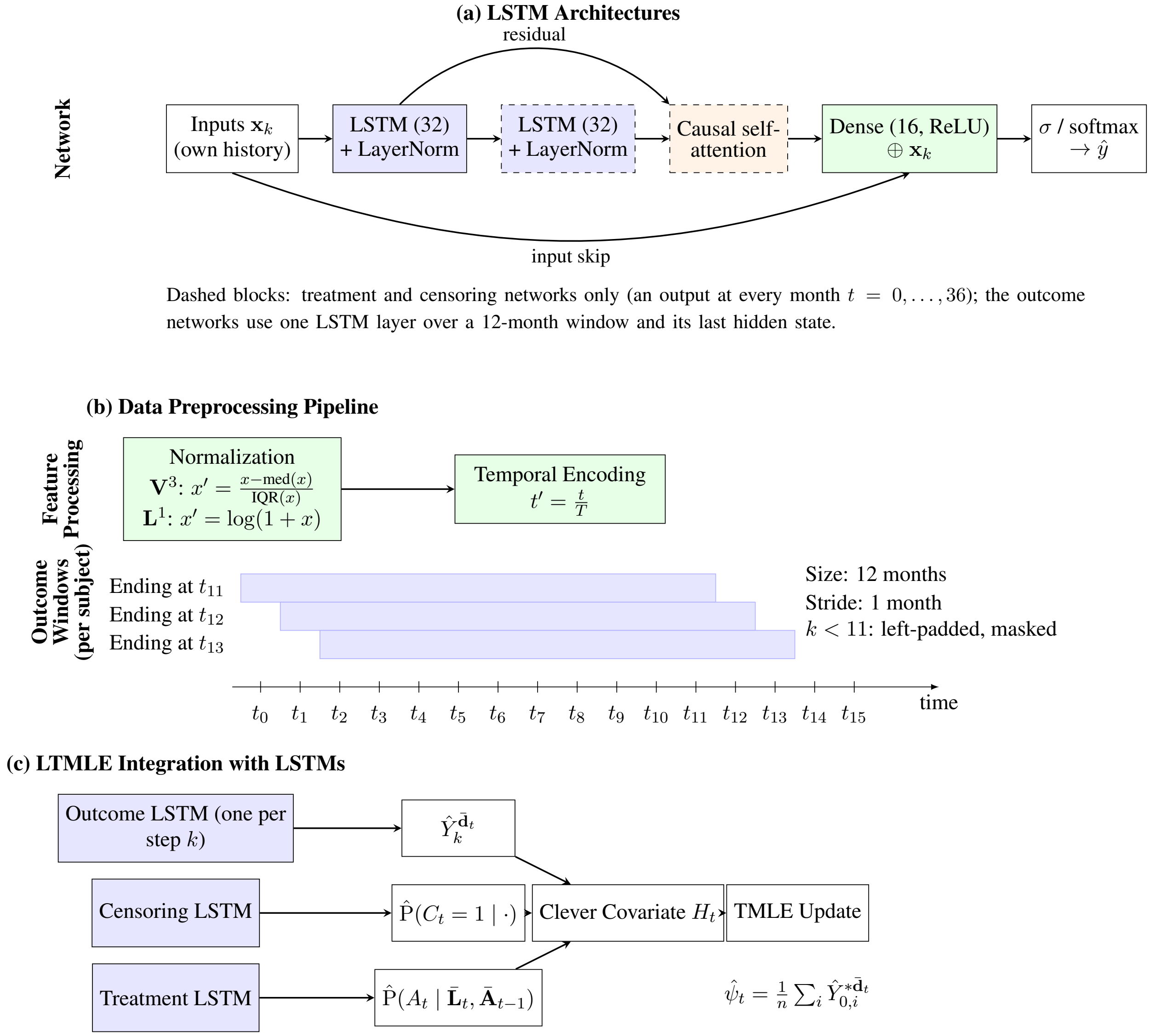

Figure \ref{fig:rnn_ltmle} illustrates our approach to integrating RNNs into the LTMLE framework. Panel (a) sketches the LSTM networks for the treatment, censoring, and outcome models, described in Section~\ref{RNN-training}. The preprocessing pipeline (panel b) shows how features are prepared through robust scaling of continuous variables (e.g., $\mathbf{V}^3$) and log-transformation of count data (e.g., $\mathbf{L}^1$). Time is encoded as $t/T$; months before baseline in an outcome window are left-padded and masked. Panel (c) shows how separate LSTM networks for outcome, censoring, and treatment predictions are combined through clever covariates to compute targeted updates of the causal parameter $\psi_t$.

\subsection{Implementation details}\label{implementation}
This section outlines the specific implementation details for the SL and RNN-based approaches.

\subsubsection{SL ensemble and training}
The SL \citep{van2007super,polley2010super,polley2011super,gruber2012tmle,phillips2023practical} addresses the challenge of model misspecification in complex longitudinal studies by employing a stacking strategy. It combines multiple prediction algorithms using a weighted linear combination, optimized to minimize $k$-fold cross-validation error. This ensemble approach leverages the strengths of diverse learners, potentially achieving better prediction accuracy than individual algorithms. By integrating various (including simple) learners into a single model, the SL reduces the risk of misspecification and effectively handles complex multivariate relationships.

We implement the SL using the \texttt{sl3} package in \texttt{R}, which supports multinomial classification algorithms and a multinomial loss function \citep{coyle2021sl3-rpkg}. Our SL ensemble includes random forests (\texttt{ranger}, \citealp{wright2017ranger}), $\ell_1$-penalized lasso regression, $\ell_1$ and $\ell_2$-penalized elastic net regression (\texttt{glmnet}, \citealp{friedman2010regularization}), a generalized linear model, and a simple mean learner (the latter two for binomial models only). Hyperparameter settings for all algorithms are provided in Table~\ref{tab:hyperparams} in the Supporting Information (SI) at the end of this article.

For multinomial treatment models, the SL combines algorithmic predictions using a linear multinomial metalearner. For binomial treatment or outcome models, the SL combines algorithmic predictions using binomial logistic regression. The SL weights are optimized using cross-validation with a negative log-likelihood loss function, where we use 3-fold cross-validation. The routine for optimizing the SL weights, given the loss function and combination function, is nonlinear optimization using Lagrange multipliers (\texttt{solnp}, \citealp{ghalanos2012rsolnp}).

\subsubsection{RNN architecture and training}\label{RNN-training}
The treatment and censoring models are many-to-many networks that output a prediction at every month: two LSTM layers of 32 units, each followed by layer normalization, then a single-head causally masked scaled dot-product self-attention module and dropout. Residual connections (labeled ``residual'' in Figure~\ref{fig:rnn_ltmle}a) add the input of the second LSTM layer and of the attention module to their outputs, so that these layers learn corrections to the representation of the first LSTM layer rather than a new one, which stabilizes the training of the stacked layers. The outcome models are sequence-to-scalar networks with one LSTM layer of 32 units over the subject's last 12 months of history, using the final hidden state, with a masking layer for left-padded months. In all networks, the output of a 16-unit ReLU pre-output layer is concatenated with the current month's inputs before the output layer (the ``input skip'' in Figure~\ref{fig:rnn_ltmle}a). The output layer is thus a logistic (for the multinomial treatment model, multinomial logistic) regression on the current month's covariates and treatment, to which the recurrent layers add learned features of the history; in particular, the outcome networks can represent the effect of setting $A_k = d_k$ directly.

The model is trained to optimize the loss function using the \texttt{AdamW} optimizer with a constant learning rate. For censoring models, we use binary cross-entropy, with censored values masked from the loss calculation. For binary outcome and treatment models, we use binary cross-entropy loss with a masking mechanism that excludes censored observations from the gradient computation. For all binary models, the output layer employs bias initialization based on the observed class ratios to accelerate training convergence, with predictions bounded to valid probability ranges. 
For multiclass treatment models, we use categorical cross-entropy loss with softmax activation, incorporating a masked loss function that ignores censored values and ensures proper probability normalization. The bias initialization for the categorical treatment model is computed as \(\text{init\_bias} = \log(\text{class\_probs} + \epsilon)\), where \(\text{class\_probs}\) are the observed class frequencies and \(\epsilon\) is a small constant for numerical stability. In all cases, model evaluation metrics are similarly masked to assess performance only on non-censored observations.

The treatment and censoring networks process each subject's own monthly sequence ($t=0,\ldots,36$) and output a prediction at every month from the history up to that month. Each outcome regression at step $k$ uses the subject's own last 12 months of history ($t_{k-11}$ to $t_k$); for $k<11$ the window is shorter, left-padded and masked, so no prediction uses information after month $k$. Early stopping uses a 10\% subject-level validation split. For the binomial specification, the RNN uses $J=6$ separate one-versus-rest networks. 

\subsubsection{Use of AI tools}\label{ai-use}
Large language model (LLM) coding agents were used in developing the software and in revising this manuscript. Claude Code (Anthropic) and Codex (OpenAI) were used to debug the \texttt{R} and \texttt{Python} simulation code. Claude Code with Claude Opus 5.5 (Anthropic) was used to audit the code against the estimators described in Section~\ref{estimators}; to rewrite parts of the SL and RNN estimation routines; to write the scripts that run the simulations and produce the tables and figures; and to update the code documentation.

\section{Simulations}\label{simulations}
In this section, we evaluate the performance of LTMLE with a single multinomial model under static, dynamic, and stochastic treatment rules, and contrast it with the same estimator using separate binary treatment models for each level. We also include comparisons of LTMLE estimated with SL versus RNNs. As singly-robust benchmarks, we include IPTW and g-computation estimators in the comparison.

The simulation process involves three primary steps. First, it generates longitudinal data capturing baseline and follow-up periods using a structural equation model. Second, it simulates observed and counterfactual outcomes for 10,000 patients who continued to follow each of three different treatment regimes: (1) \emph{static}: patients remain on their initial medication; (2) \emph{dynamic}: patients switch medications if they experience metabolic complications or acute mental health events; and (3) \emph{stochastic}: at each timepoint, patients have a 95\% probability of maintaining their current medication and 5\% probability of switching according to a multinomial distribution. This process is repeated for 100 simulation runs. 

Third, the process applies each estimator to calculate the counterfactual outcome at months $t=12$, 24, and 36 under no censoring. All initial outcome models are estimated using either SL or an RNN with appropriate activation functions and loss functions for binary outcomes. We calculate four performance metrics (formalized in the Supporting Information): (i) absolute bias, or the average absolute difference between estimated and true parameter values; (ii) coverage probability, or the proportion of estimated 95\% confidence intervals (CIs) containing the true parameter value; (iii) the average width of estimated CIs; and (iv) the proportion of uncensored rule-followers whose estimated cumulative probability of following the assigned rule through time $t$ is small ($<0.025$), which measures data support for a given intervention rule and serves as a diagnostic tool for detecting positivity violations \citep{schomaker2019using}.

\subsection{Generated data}

We generate longitudinal data ($t = 0, 1, \ldots, 36$) for time-invariant baseline variables and time-varying variables using structural equations implemented with the \texttt{simcausal} package in \texttt{R} \citep{sofrygin2018simcausal}. The data generating process (DGP) is formalized in Appendix A.1 and illustrated by a directed acyclic graph (DAG) in Figure \ref{fig:DAG}, which visualizes complex temporal dependencies across the first four time points. At baseline ($t = 0$), we simulate race ($\mathbf{V}^1$, categorical), primary SMI diagnosis ($\mathbf{V}^2$, categorical, with probabilities varying by race), and age ($\mathbf{V}^3$, normal with mean and standard deviation varying by SMI diagnosis, bounded between 19.9 and 64.5 years). Diagnosis ($\mathbf{V}^2$) influences the initial values of time-varying covariates --- psychiatric emergency department visits ($\mathbf{L}^1_t$, negative binomial), metabolic testing ($\mathbf{L}^2_t$, binary), and antidiabetic medication use ($\mathbf{L}^3_t$, binary) --- and treatment ($A_t$, a 6-level categorical variable representing aripiprazole, haloperidol, olanzapine, quetiapine, risperidone, and ziprasidone, with probabilities depending on SMI diagnosis and time-varying covariates); age affects only censoring. Additional time-varying variables include a censoring indicator ($C_t$, binary), which occurs deterministically when patients reach age 65 to reflect Medicare eligibility and stochastically based on covariate and treatment history, and a diabetes diagnosis outcome ($Y_t$, binary), with probability determined through a logistic model incorporating previous diagnosis status, changes in ED visits, metabolic testing, antidiabetic medication use, and current and previous treatment. For subsequent periods ($t > 0$), each variable depends on variables at times $t$ and $t-1$ (a first-order Markov structure) --- for example, psychiatric ED visits ($\mathbf{L}^1_t$) are influenced by previous visits, metabolic testing ($\mathbf{L}^2_{t-1}$), antidiabetic prescriptions ($\mathbf{L}^3_{t-1}$), and the previous treatment ($A_{t-1}$) --- while treatment decisions ($A_t$) rely on current time-varying covariates and the previous treatment, and the outcome ($Y_t$) and censoring ($C_t$) processes depend on current and previous covariates and treatments. The generated data mimics the patient-level information collected by the Centers for Medicare \& Medicaid Services (CMS), which is used in the studies of \citet{poulos2023antipsychotics,poulos2024targeted}.

\subsection{Causal parameters}
We define causal parameters of interest, $\psi_{j,t}$, as the expected outcome value at each time point $t$, under specified treatment rules $\bar{\mathbf{d}}^j$, and in the absence of censoring. Below, we formalize the three treatment rules --- static, dynamic, and stochastic--- in the context of the simulation study. According to the static rule, patients receive a specific medication based on their primary SMI diagnosis on $t=0$ and stay on it consistently:
\begin{flalign*}
\bar{\mathbf{d}}_{t}^{1}  = \begin{cases}
A_t = \text{quet.}, & \text{if bipolar } (\mathbf{V}^2=2); \\
A_t = \text{halo.}, & \text{if schizophrenia } (\mathbf{V}^2=3); \\
A_t = \text{ari.}, & \text{if MDD } (\mathbf{V}^2=1). \\
\end{cases}
\end{flalign*}

\noindent
Under the dynamic rule, everyone starts with risperidone. For $t>0$, if a psychiatric ED visit, metabolic testing, or an antidiabetic prescription fill is observed ($\mathbf{L}^1_t > 0$ or $\mathbf{L}^2_t > 0$ or $\mathbf{L}^3_t > 0$), then patients switch to a condition-specific medication; otherwise, they stay on (or return to) risperidone:
\begin{flalign*}
\bar{\mathbf{d}}_{t}^{2} = \begin{cases}
A_0 = \text{risp.}; \\
A_t = \text{quet.}, & \text{if bipolar } (\mathbf{V}^2=2) \text{ and } (\mathbf{L}^1_t > 0 \text{ or } \mathbf{L}^2_t > 0 \text{ or } \mathbf{L}^3_t > 0); \\
A_t = \text{halo.}, & \text{if schizophrenia } (\mathbf{V}^2=3) \text{ and } (\mathbf{L}^1_t > 0 \text{ or } \mathbf{L}^2_t > 0 \text{ or } \mathbf{L}^3_t > 0); \\
A_t = \text{ari.}, & \text{if MDD } (\mathbf{V}^2=1) \text{ and } (\mathbf{L}^1_t > 0 \text{ or } \mathbf{L}^2_t > 0 \text{ or } \mathbf{L}^3_t > 0); \\
A_t = \text{risp.}, & \text{otherwise}.
\end{cases}
\end{flalign*}
\noindent
Under the stochastic rule, for $t>0$, there is a 95\% probability of staying with treatment at $t-1$, and a 5\% probability of randomly switching according to a multinomial distribution:
\begin{flalign*}
\Pr(A_t=a \mid A_{t-1}=a_{t-1}) = \begin{cases}
0.95, & \text{if } a=a_{t-1}; \\
\frac{0.05}{J-1}, & \text{if } a\neq a_{t-1},
\end{cases}
\end{flalign*}
where $J=6$ is the number of treatment options; baseline treatment is not intervened upon.

The true values of these causal parameters are obtained through simulation using the \texttt{simcausal} package, which generates counterfactual data under each treatment rule with no censoring. These parameters represent the cumulative incidence of diabetes diagnosis over time under different treatment strategies. Comparing causal parameters under different treatment rules allows for an evaluation of the expected outcomes associated with each rule. Through this comparison, it's possible to identify which treatment rules are most likely to lead to a reduction in patient risks. For instance, a static rule may suggest stability in treatment effects, but might not adapt to changes in patient condition. The dynamic rule, with its adjustments based on patient response can lead to more targeted care, but may introduce variability that complicates the estimation of causal effects. Finally, the stochastic rule introduces randomness that could capture potential benefits of unanticipated treatment changes.

Figure~\ref{fig:combined_simulation} provides insights into the DGP through a comparative analysis of treatment adherence and survival probability under the treatment rules in a single simulation run. Panel (a) shows the proportion of uncensored patients who continued to follow the dynamic and static rules over time; the stochastic rule assigns a positive probability to every treatment path, so all uncensored patients remain compatible with it. The dynamic and static rules start with adherence of about 17 to 18\%, respectively, and show a gradual decline throughout the study period, reaching about 9 to 10\% by month 12 and 3 to 4\% by month 36.

Panel (b) displays the observed survival curves, or the proportion of patients without a diabetes diagnosis, under each treatment rule in the presence of censoring. The dynamic rule shows the highest survival probability (about 78\% at month 36), followed by the stochastic rule (71\%, i.e., all uncensored patients), while the static rule results in the lowest survival (64\%). However, these observed differences are confounded by both censoring and differential treatment adherence.

Panel (c) presents the true counterfactual survival curves that would be observed under perfect treatment adherence and no censoring. The ordering differs from the observed data: the dynamic rule maintains the highest survival probability (about 93\% at month 36), followed by the static rule (85\%), while the stochastic rule shows the lowest survival (71\%), similar to its observed value. These differences between observed and counterfactual outcomes highlight the importance of properly adjusting for censoring and treatment adherence when estimating causal effects.

\subsection{Estimators}
We evaluate multiple LTMLE implementations that differ in their treatment model specification and estimation approach. The first set of estimators implements SL for estimating the outcome, treatment, and censoring probabilities: LTMLE using multinomial treatment probabilities, LTMLE-SL (multi.); or, using separate binary treatment models, LTMLE-SL (bin.). In the binomial specification, $J$ separate one-versus-rest models $\hat{p}_{k,j} = \hat{\Pr}(A_k = j \mid \bar{\mathbf{L}}_k, \bar{\mathbf{A}}_{k-1}, \bar{\mathbf{C}}_{k-1}=1)$, $j = 1, \ldots, J$, are fit at each time, and $\hat{p}_{k,d_k}$ enters $\hat{g}_t$ and $H_t$ without renormalization, so the $\hat{p}_{k,j}$ need not sum to one. For comparison, we include IPTW estimators with both multinomial, IPTW-SL (multi.) and binary treatment models, IPTW-SL (bin.). We also compare with g-computation, G-Comp-SL.

The second set of estimators replaces SL with RNNs in the form of LSTM networks: LTMLE-RNN with both multinomial and binary treatment models, IPTW-RNN with both treatment model specifications, and G-Comp-RNN.

The implementation details for both the SL and RNN implementations, including model architectures, training procedures, and hyperparameter settings, are described in Section \ref{implementation}. LTMLE with multinomial treatments is theoretically preferred as it correctly specifies the multinomial structure of treatment probabilities and maintains the double robustness property. LTMLE with binomial treatments represents a common but potentially misspecified approach that treats each treatment level as a separate binary decision. The IPTW and g-computation estimators serve as singly-robust benchmarks, while the RNN-based estimators explore whether deep learning approaches can better capture temporal dependencies in the data compared to traditional statistical learning methods.

\subsection{Results}\label{results}
Our simulation results reveal distinct patterns in estimator performance across different treatment rules and over time. For SL-based implementations, both LTMLE-SL (multi.) and LTMLE-SL (bin.) are approximately unbiased, with mean absolute bias of 0.006 and mean bias no larger than 0.0025 in absolute value at any rule and time point, and achieve coverage close to the nominal level (0.940 and 0.934; Table \ref{tab:results-estimator}). Coverage of LTMLE-SL (multi.) ranges from 0.920 under the dynamic rule to 0.973 under the static rule (Table~\ref{tab:results-estimator-rule}), where its intervals are conservative at $t=36$ (Figure~\ref{fig:coverage}).

The IPTW-SL estimators perform similarly to LTMLE-SL under the dynamic rule but are biased at later time points under the static and stochastic rules (Figure~\ref{fig:bias}). IPTW-SL (multi.) underestimates the risk at $t=36$ under the static and stochastic rules, with coverage of 0.70 and 0.43, whereas IPTW-SL (bin.) is less biased (coverage 0.76 and 0.97; Figure~\ref{fig:coverage}). This result is consistent with the multinomial SL producing less well-calibrated small treatment probabilities, to which IPTW is sensitive but LTMLE is not. IPTW depends on the treatment model only through the inverse of the estimated cumulative probability of following the rule, a product of up to 37 monthly probabilities, and at $t=36$ most rule-followers have cumulative probabilities below 0.025 (Table~\ref{tab:positivity}). Small systematic errors in the monthly probabilities therefore compound into large errors in the weights of the patients who determine the estimate. A plausible source of such errors is that the multinomial SL library, unlike the binomial library, contains no unpenalized logistic regression, so the lasso and elastic-net penalties shrink the predicted probabilities of rarely received treatments toward the marginal treatment frequencies. LTMLE uses the same weights only in its targeting step, and its bias depends on the product of the errors of the treatment and outcome models, so the outcome regressions largely correct errors in the treatment model. G-Comp-SL is substantially biased under the dynamic rule, overestimating the risk by 0.018, 0.037, and 0.059 at $t=12$, 24, and 36, so that its approximate intervals rarely contain the truth. To diagnose this bias, we repeated the g-computation recursion for the dynamic rule at $t=36$ in one simulated dataset, once with SL and once with main-terms logistic regressions as the outcome models. The SL estimate was 0.122 and the logistic-regression estimate 0.065 (true value 0.068). At steps $k<t$ the pseudo-outcomes are probabilities rather than binary outcomes, and the SL combines least-squares versions of its learners with a linear metalearner; its ensemble weights, averaged over these steps, were concentrated on the lasso, elastic-net, and linear regressions (0.36, 0.28, and 0.22, with 0.07 each on the mean and random forests). These regressions shrank the predicted change in risk from setting each month's treatment to 71\% of that under the logistic regressions. Since only 17 to 25\% of the patients at risk receive the rule's treatment in a given month, setting treatment to the rule changes the treatment of most patients, and the shrinkage compounds over the 37 backward steps, pulling the estimate toward the natural-course risk; that is, the risk under the treatments patients actually received without intervention, which is higher than under the dynamic rule.

When examining performance by treatment rule (Table \ref{tab:results-rule}), we aggregate results separately across SL and RNN implementations to assess systematic differences in estimation difficulty. For the SL implementations, which combine results from LTMLE-SL, G-Comp-SL, and IPTW-SL estimators, we find that estimation under the stochastic rule achieves the smallest mean absolute bias (0.005) and the narrowest confidence intervals (median width 0.018). The dynamic rule presents the greatest challenge, with the lowest coverage probability (0.743), driven by G-Comp-SL, and the highest mean absolute bias (0.013). The static rule yields coverage of 0.907 and the widest intervals (median width 0.047). For the RNN implementations, which aggregate across LTMLE-RNN, G-Comp-RNN, and IPTW-RNN estimators, coverage is 0.953, 0.918, and 0.857 under the static, dynamic, and stochastic rules, with mean absolute bias of 0.009, 0.008, and 0.005, respectively.

Figure~\ref{fig:sim_results_long} illustrates how these patterns evolve over time. Panel (a) shows that absolute bias increases over time for most estimators, most steeply for G-Comp-SL and for the IPTW estimators under the static rule, while the LTMLE estimators remain close to unbiased at every time point. G-Comp-RNN shows the smallest bias under the static and dynamic rules but, like G-Comp-SL, increasingly underestimates the risk under the stochastic rule.

Panel (b) shows that LTMLE-SL (multi.) maintains coverage near the nominal level throughout the study period, whereas the coverage of LTMLE-RNN declines to 0.82--0.87 at $t=36$ under the dynamic and stochastic rules, where its standard errors underestimate the empirical standard deviation by up to 26\%. Since the networks are fit and evaluated on the same subjects without cross-fitting, in-sample overfitting can shrink the residuals in \eqref{eq:IC} --- a plausible explanation for this underestimation, which also occurs under the stochastic rule without positivity problems. IPTW-SL (multi.) under-covers at later time points under the static and stochastic rules, and both g-computation estimators under-cover under the stochastic rule at $t=36$.
Panel (c) reveals gradually widening confidence intervals over time, particularly pronounced for the static rule. LTMLE-RNN produces narrower intervals than LTMLE-SL under the static rule (0.062 at $t=36$), where the SL standard errors are conservative.

The RNN-based implementations perform similarly to the SL-based implementations for LTMLE: LTMLE-RNN (multi.) has the same mean absolute bias as LTMLE-SL (multi.) and slightly lower coverage, so RNNs do not reduce the bias of LTMLE relative to SL in this setting. Unlike G-Comp-SL, however, G-Comp-RNN is approximately unbiased under the dynamic rule, suggesting that the RNN outcome regressions do not attenuate the treatment contrast in the same way. The binomial RNN treatment models are less stable than the multinomial RNN model: in one run they assigned very small probabilities to the rule-prescribed treatment for many patients, inflating the standard errors of LTMLE-RNN (bin.), and IPTW-RNN (bin.) has 6 to 8 times the variance of IPTW-RNN (multi.) under the stochastic rule at $t=24$ and 36.

Looking across all metrics and scenarios, several key findings emerge. First, the choice between multinomial and binomial treatment models is inconsequential for LTMLE-SL but matters for IPTW and for the RNN implementation, where the multinomial model yields more stable inference. Second, LTMLE maintains consistent performance across treatment rules with either SL or RNNs, whereas IPTW and g-computation are sensitive to the treatment and outcome learners, respectively. Third, the dynamic rule is the most challenging for g-computation with SL, and the coverage of most estimators is lowest at $t=36$, when the fewest patients follow the static and dynamic rules.

\begin{table}[ht]
\centering
\caption{Performance metrics by estimator, aggregated across time points and treatment rules. Values are over 100 simulation runs ($n=10{,}000$) and target time points $t=12, 24, 36$. Coverage probability: proportion of 95\% CIs containing the true value (Monte Carlo SE); CI widths: mean (median); absolute bias: mean $\pm$ SD over all replicate-by-cell values. G-computation intervals are approximate (influence curve evaluated at the untargeted fits).}
\label{tab:results-estimator}
\resizebox{0.8\textwidth}{!}{%
\begin{tabular}{l|c|c|c}
\hline
\textbf{Estimator} & \textbf{Coverage prob.} & \textbf{CI Widths} & \textbf{Abs. Bias} \\
\hline
G-Comp-SL & $0.588$ (0.016) & $0.0352$ (0.0273) & $0.0208 \pm 0.0178$ \\ 
IPTW-SL (bin.) & $0.917$ (0.009) & $0.0324$ (0.0251) & $0.0075 \pm 0.0086$ \\ 
IPTW-SL (multi.) & $0.807$ (0.013) & $0.0322$ (0.0248) & $0.0092 \pm 0.0094$ \\ 
LTMLE-SL (bin.) & $0.934$ (0.008) & $0.0329$ (0.0266) & $0.0063 \pm 0.0067$ \\ 
LTMLE-SL (multi.) & $0.940$ (0.008) & $0.0344$ (0.0272) & $0.0064 \pm 0.0068$ \\ 
\hline
G-Comp-RNN & $0.929$ (0.009) & $0.0299$ (0.0249) & $0.0054 \pm 0.0042$ \\ 
IPTW-RNN (bin.) & $0.899$ (0.010) & $0.0343$ (0.0260) & $0.0087 \pm 0.0118$ \\ 
IPTW-RNN (multi.) & $0.901$ (0.010) & $0.0341$ (0.0259) & $0.0084 \pm 0.0090$ \\ 
LTMLE-RNN (bin.) & $0.906$ (0.010) & $0.0499$ (0.0257) & $0.0067 \pm 0.0069$ \\ 
LTMLE-RNN (multi.) & $0.913$ (0.009) & $0.0297$ (0.0248) & $0.0064 \pm 0.0066$ \\ 
\hline
\end{tabular}
}
\end{table}

\begin{table}[ht]
\centering
\caption{Performance metrics by treatment rule, aggregated across estimators (LTMLE, IPTW, and G-computation) and time points, separately by implementation. Values are over 100 simulation runs ($n=10{,}000$) and target time points $t=12, 24, 36$. Coverage probability: proportion of 95\% CIs containing the true value (Monte Carlo SE); CI widths: mean (median); absolute bias: mean $\pm$ SD over all replicate-by-cell values. G-computation intervals are approximate (influence curve evaluated at the untargeted fits).}
\label{tab:results-rule}
\resizebox{0.8\textwidth}{!}{%
\begin{tabular}{ll|c|c|c}
\hline
\textbf{Implementation} & \textbf{Rule} & \textbf{Coverage prob.} & \textbf{CI Widths} & \textbf{Abs. Bias} \\
\hline
\multirow{3}{*}{SL} & Static & $0.907$ (0.008) & $0.0494$ (0.0469) & $0.0123 \pm 0.0103$ \\ 
 & Dynamic & $0.743$ (0.011) & $0.0331$ (0.0308) & $0.0130 \pm 0.0163$ \\ 
 & Stochastic & $0.861$ (0.009) & $0.0177$ (0.0181) & $0.0048 \pm 0.0041$ \\ 
\hline
\multirow{3}{*}{RNN} & Static & $0.953$ (0.005) & $0.0567$ (0.0441) & $0.0089 \pm 0.0085$ \\ 
 & Dynamic & $0.918$ (0.007) & $0.0322$ (0.0301) & $0.0076 \pm 0.0086$ \\ 
 & Stochastic & $0.857$ (0.009) & $0.0178$ (0.0179) & $0.0049 \pm 0.0068$ \\ 
\hline
\end{tabular}
}
\end{table}

To measure the data support for a given intervention rule, we calculate, among uncensored rule-followers, \(\prod_{s=0}^{t} \hat{P}(A_s = d_s \mid \bar{\mathbf{L}}_s = \bar{\mathbf{l}}_s, \bar{\mathbf{A}}_{s-1} = \bar{\mathbf{d}}_{s-1})\), i.e. the estimated cumulative probability of continuing to receive treatment according to the assigned treatment rule through time $t$, given the covariate history (for the stochastic rule, the cumulative ratio of the estimated to the rule's treatment probabilities). A substantial proportion of units that have small cumulative probabilities (i.e., $< 0.025$) is indicative of positivity violations \citep{schomaker2019using}. Table~\ref{tab:positivity} reports the measure per intervention rule and selected time period: under the static and dynamic rules, almost no rule-followers have small cumulative probabilities at $t=12$ and 24, but at $t=36$ the proportion is 0.93 and 0.95 with SL and 0.38 and 0.27 with RNNs, whereas the stochastic rule shows no evidence of positivity violations.

Table~\ref{tab:run-times} reports the computation time of the simulation runs, each with three worker processes on a laptop CPU without a GPU. A run took a median of 25 minutes with the RNN implementation and 56 minutes with SL, so the RNNs were about 2.2 times faster. The two implementations spend their time differently. With SL, most of the time (39 minutes) goes to the treatment and censoring models, for which separate SLs are fit at each of the 37 time points (a multinomial model, six binomial models, and a censoring model). With RNNs, each treatment and censoring model is a single LSTM fit to the whole sequence (4.5 minutes in total), and most of the time (21 minutes) goes to the outcome regressions, which fit an LSTM at every step of the backward recursion.

\section{Discussion}
Our simulation results reveal several important insights about the performance of LTMLE estimators with multi-valued treatments under different treatment rules. First, the distinction between multinomial and binomial treatment model specifications is inconsequential for LTMLE with SL, but the multinomial model yields more stable inference when the treatment models are RNNs, and the specification affects IPTW with either learner. This suggests that the treatment model specification matters mainly for estimators that rely solely on the estimated weights, and that a multinomial model is preferable when the treatment mechanism is estimated with RNNs.

Second, estimator performance varies by treatment rule. The dynamic rule, despite showing the best survival outcomes, presents the greatest challenge for g-computation with SL, which is biased by up to 0.059 because the rule departs most from observed practice and few patients follow it, so attenuated treatment contrasts accumulate over the backward recursion. LTMLE corrects this bias through targeting, and its performance under the dynamic rule is comparable to that under the static and stochastic rules, apart from somewhat lower coverage at later time points when few patients remain on the rule; LTMLE-RNN also under-covers at $t=36$ under the stochastic rule, consistent with optimistic influence-curve variance estimates from neural networks fit without cross-fitting.

Third, the low treatment rule adherence observed in our simulation mirrors real-world challenges in maintaining consistent treatment regimes. The consistently low adherence to the static and dynamic rules (about 17 to 18\% at baseline and 3 to 4\% by month 36) underscores the importance of methods that can handle substantial treatment deviations. In our simulations, IPTW estimators, which rely solely on the resulting weights, are more sensitive to the treatment model than LTMLE, which also uses the outcome regressions.

\section{Conclusion}
This study examines LTMLE with multi-valued treatments in longitudinal settings. Our results show that LTMLE with either SL or RNNs is approximately unbiased with close to nominal coverage, that binary and multinomial treatment models perform similarly for LTMLE with SL while the multinomial model gives more stable inference with RNNs, and that plug-in g-computation under dynamic rules can be badly biased. These findings inform the implementation of causal inference methods with multiple treatment options.
Extensions to this work should address two key issues. First, positivity violations arise when subgroups rarely receive certain treatments, particularly with rarely followed treatment regimes. These violations occur both theoretically --- when it is structurally impossible for individuals with specific covariate values to receive certain exposures --- and practically, through data sparsity in finite samples. This problem persists regardless of sample size and worsens with continuous treatments or extensive covariate sets. Solutions include designing interventions that match observed treatment patterns \citep{hoffman2023introducing}, truncating density ratios at specific thresholds during exploratory analysis, and developing methods to detect and handle violations \citep{schomaker2024causal,mcclean2026comparing}. However, approaches like density ratio truncation may introduce their own biases.

Second, the traditional TMLE framework treats propensity score estimation as an independent statistical task, which can lead to suboptimal bias-variance trade-offs in the estimation of the target parameter. This limitation becomes particularly evident in small sample settings, where overfitting or near-perfect prediction in the propensity score model can exacerbate positivity violations and introduce numerical instability. An end-to-end TMLE approach using a unified neural network architecture could address these challenges by jointly optimizing all components of the procedure, building on recent RNN- and transformer-based approaches \citep{frauen2023estimating,shirakawa2024longitudinal}; cross-fitting (CV-TMLE) of the neural network nuisance models is a further natural extension. This would ensure that propensity score estimation is directly aligned with the goal of minimizing bias and variance in the target parameter estimation, rather than being optimized independently. The unified architecture would allow for shared feature representations across different components of the estimator and enable automatic balancing of the bias-variance trade-off through end-to-end training. This approach is particularly promising in small sample settings where traditional TMLE often struggles with positivity violations, as the joint optimization can help prevent extreme propensity score estimates that lead to instability in the targeting step.

These extensions, combined with our findings about estimator behavior under different rules, outline next steps for causal inference in longitudinal settings. The challenges identified emphasize the need for careful study design, particularly regarding positivity violations and treatment rule specification, when implementing these methods in practice \citep{nance2024causal}. The development of end-to-end approaches may provide more robust solutions for real-world applications where sample sizes are limited and traditional sequential estimation procedures may be suboptimal.

\vspace*{1pc}

\noindent {\bf{Acknowledgements}}

\noindent The author acknowledges the use of Claude Code (Anthropic) with Claude Opus 5.5 to revise and draft manuscript text and to format the LaTeX source, and of Claude Code and Codex (OpenAI) for code development, as described in Section~\ref{ai-use}. The author designed the study; specified the estimators, data-generating process, and analyses; directed the AI tools; reviewed and edited all AI-generated code and text; and checked the simulation results. The author takes full responsibility for the content of this article.

\vspace*{1pc}

\noindent {\bf{Conflict of interest statement}}

\noindent The author declares no conflict of interest.

\vspace*{1pc}

\noindent {\bf{Data availability statement}}
\noindent 
The code that generates the simulated data and reproduces all results in this article is openly available at \url{https://github.com/jvpoulos/multi-ltmle}. The repository contains the \texttt{R} and \texttt{Python} scripts for data generation, estimation, the simulation study, and the tables and figures; all simulation parameters and random seeds are set in the code.

\section*{Appendix}
\setcounter{figure}{0}
\renewcommand{\thefigure}{A\arabic{figure}}

\subsection*{A.1.\enspace Simulation DGP}\label{DGP}
The distributions below describe the data-generating process for the baseline data ($t = 0$) and time-varying data ($t = 1,\ldots,36$). Baseline data consists of race, primary diagnosis, age, psychiatric ED visits, metabolic testing, and antidiabetic prescription fill ($\mathbf{V}^1$, $\mathbf{V}^2$, $\mathbf{V}^3$, $\mathbf{L}^1_0$, $\mathbf{L}^2_0$, $\mathbf{L}^3_0$, respectively). Time-varying data consists of psychiatric ED visits, metabolic testing, antidiabetic prescription fills, outcome, treatment, and censoring ($\mathbf{L}^1_t$, $\mathbf{L}^2_t$, $\mathbf{L}^3_t$, $Y_t$, $A_t$, $C_t$).

The Bernoulli ($\text{B}(p)$) and generalized Bernoulli ($\text{GB}(p_1, \ldots, p_k)$) distributions are used to generate binary and categorical variables, respectively. The negative binomial ($\text{NegBinom}(\mu)$) distribution with mean $\mu$ and size (dispersion) parameter 1 is used to generate ED visits, and the normal distribution $\mathcal{N}_{[a, b]}(\mu, \sigma^2)$ with mean $\mu$ and variance $\sigma^2$, with draws below $a$ or above $b$ replaced by uniform draws on $[19.9, 36.79]$ or $[53.27, 64.5]$, respectively, is used to generate age. The multinomial distribution\ $\text{Multinom}(p_1, \ldots, p_k)$ is used for the categorical treatment.

\begin{figure}[htbp]
\centering
\includegraphics[width=\textwidth]{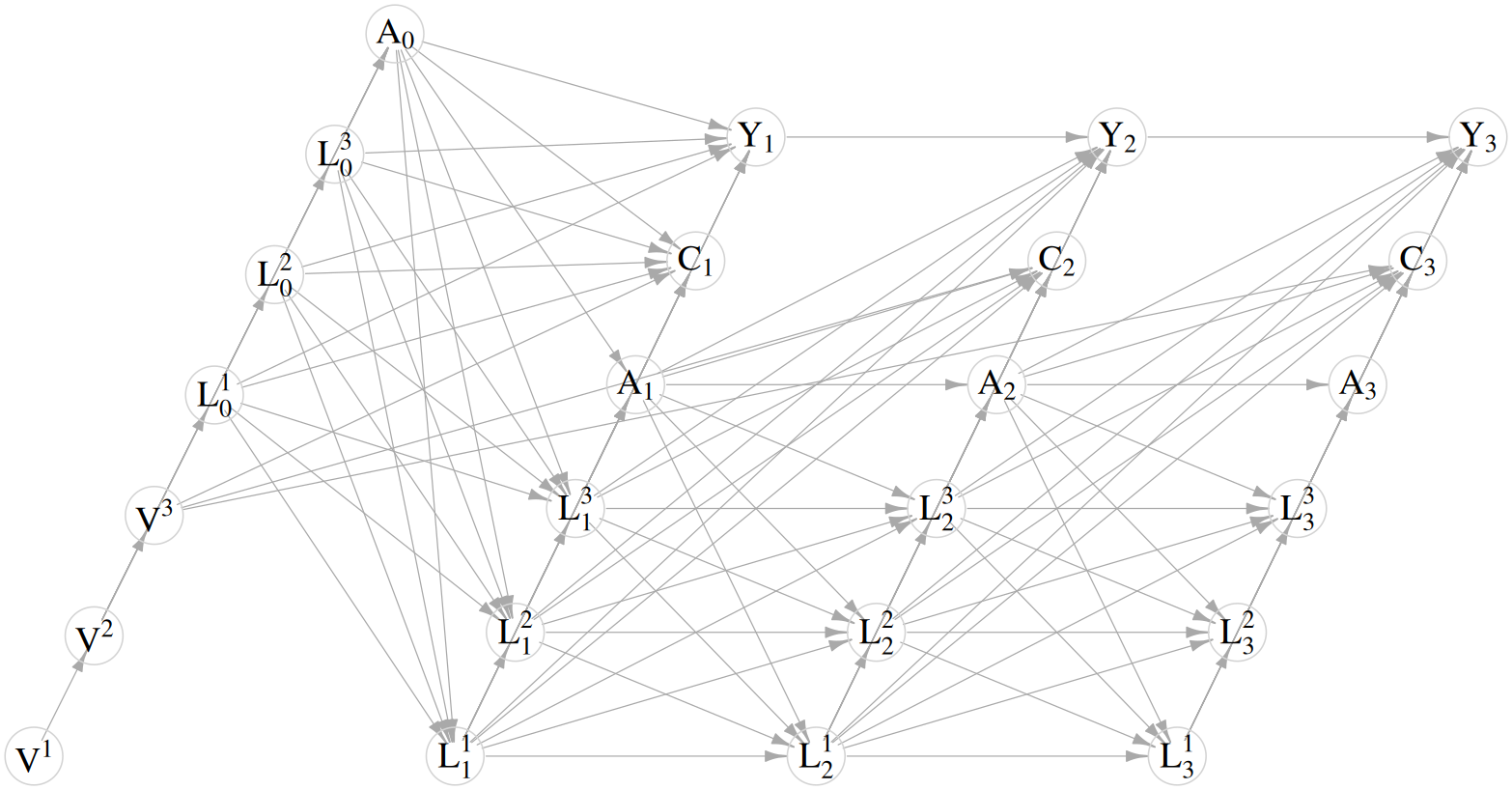}
\caption{Directed acyclic graph (DAG) illustrating the temporal relationships between baseline characteristics ($\mathbf{V}^1$, $\mathbf{V}^2$, $\mathbf{V}^3$), psychiatric ED visits ($\mathbf{L}^1_t$), metabolic testing ($\mathbf{L}^2_t$), antidiabetic prescriptions ($\mathbf{L}^3_t$), treatment ($A_t$), censoring ($C_t$), and diabetes diagnosis ($Y_t$) for $t = 0,1,2,3$ ($C_0$ and $Y_0$ are degenerate and omitted). Arrows represent direct causal effects.}
\label{fig:DAG}
\end{figure}

\subsubsection*{A.1.1.\enspace Baseline and time-varying covariates}
At baseline (t = 0), the covariates are generated as follows:
\begin{flalign*}
&\textbf{Race (White, Black, Latino, Other)}: \mathbf{V}^1 \sim \text{GB}(0.65, 0.17, 0.11, 0.07), &\\
&\textbf{SMI condition (MDD, Bipolar, Schizophr.)}: \mathbf{V}^2 \mid \mathbf{V}^1 \sim
\begin{cases}
\text{GB}(0.17, 0.20, 0.63), & \text{if } \mathbf{V}^1 = \text{White}, \\
\text{GB}(0.23, 0.30, 0.47), & \text{if } \mathbf{V}^1 = \text{Black}, \\
\text{GB}(0.21, 0.28, 0.51), & \text{if } \mathbf{V}^1 \in \{\text{Latino}, \text{Other}\},
\end{cases} &\\
&\textbf{Age}: \mathbf{V}^3 \mid \mathbf{V}^2 \sim \mathcal{N}_{[19.9, 64.5]}(\mu_{\mathbf{V}^2}, \sigma^2_{\mathbf{V}^2}), \text{ where}&\\
&(\mu_{\mathbf{V}^2}, \sigma^2_{\mathbf{V}^2}) =
\begin{cases}
(44.0, 10.5^2), & \text{if } \mathbf{V}^2 = \text{MDD}, \\
(44.5, 10.2^2), & \text{if } \mathbf{V}^2 = \text{Bipolar}, \\
(46.5, 9.2^2), & \text{if } \mathbf{V}^2 = \text{Schizophr.}
\end{cases}&
\end{flalign*}
Time-varying covariates at (t = 0) are generated as follows:
\begin{flalign*}
\textbf{Psychiatric ED visits}: \mathbf{L}^1_0 \mid \mathbf{V}^2 &\sim
\begin{cases}
\text{NegBinom}(0.09), & \text{if } \mathbf{V}^2 = \text{Schizophr.}, \\
\text{NegBinom}(0.05), & \text{if } \mathbf{V}^2 = \text{Bipolar}, \\
\text{NegBinom}(0.03), & \text{if } \mathbf{V}^2 = \text{MDD},
\end{cases} \\
\textbf{Metabolic testing}: \mathbf{L}^2_0 \mid \mathbf{V}^2 &\sim
\begin{cases}
\text{Bernoulli}(0.075), & \text{if } \mathbf{V}^2 = \text{Schizophr.}, \\
\text{Bernoulli}(0.05), & \text{if } \mathbf{V}^2 = \text{Bipolar}, \\
\text{Bernoulli}(0.025), & \text{if } \mathbf{V}^2 = \text{MDD},
\end{cases} \\
\textbf{Antidiabetic Rx}: \mathbf{L}^3_0 \mid \mathbf{V}^2 &\sim
\begin{cases}
\text{Bernoulli}(0.085), & \text{if } \mathbf{V}^2 = \text{Schizophr.}, \\
\text{Bernoulli}(0.015), & \text{if } \mathbf{V}^2 = \text{Bipolar}, \\
\text{Bernoulli}(0.035), & \text{if } \mathbf{V}^2 = \text{MDD}.
\end{cases}
\end{flalign*}
The time-varying covariates at $t>0$ are generated based on the conditional set $\bar{\mathcal{D}}_t = \{\mathbf{V}^1, \mathbf{V}^2, \mathbf{V}^3,$ $\bar{\mathbf{L}}_{t-1}, \bar{\mathbf{A}}_{t-1}, \bar{Y}_{t-1}\}$, which includes time-invariant baseline variables, and the histories of covariates, treatments, and outcomes up to time $t-1$, together with any variables preceding the node at time $t$ in the ordering $\mathbf{L}^1_t,\mathbf{L}^2_t,\mathbf{L}^3_t,A_t,C_t,Y_t$. The equations are defined as follows:
\begin{itemize}[label={}]
    \item \textbf{Psychiatric ED visits:}
    \begin{flalign*}
        &\mathbf{L}^1_t \mid \bar{\mathcal{D}}_t \sim \text{NegBinom}\left(\operatorname{logit}^{-1}\left(0.05 \cdot (L^1_{t-1})^2 + 0.05 \cdot L^2_{t-1} + 0.1 \cdot L^3_{t-1} + \alpha(A_{t-1})\right)\right), &\\
        &\alpha(A_{t-1}) = 
        \begin{cases}
            -1, & \text{if } A_{t-1} \in \{\text{ari., halo., quet.}\}, \\
            -5, & \text{if } A_{t-1} = \text{risp.}, \\
            0, & \text{otherwise}.
        \end{cases} &
    \end{flalign*}
    \item \textbf{Metabolic testing}
    \begin{flalign*}
        &\mathbf{L}^2_t \mid \bar{\mathcal{D}}_t \sim \text{Bernoulli}\left(p_{\mathbf{L}^2_t}\right), &\\
        &p_{\mathbf{L}^2_t} = 
            \operatorname{logit}^{-1}\left(-2 + 0.05 \cdot (\mathbf{L}^1_t - L^1_{t-1})^2 + 0.1 \cdot L^2_{t-1} + 0.1 \cdot L^3_{t-1} + \delta(A_{t-1})\right). &\\
        &\text{where } \delta(A_{t-1}) = 
        \begin{cases}
            -4, & \text{if } A_{t-1} \in \{\text{ari., halo., quet.}\}, \\
            -5, & \text{if } A_{t-1} = \text{risp.}, \\
            0, & \text{otherwise}.
        \end{cases} &
    \end{flalign*}
    \item \textbf{Antidiabetic Rx:}
    \begin{flalign*}
        &\mathbf{L}^3_t \mid \bar{\mathcal{D}}_t \sim \text{Bernoulli}\left(p_{\mathbf{L}^3_t}\right), &\\
        &p_{\mathbf{L}^3_t} = 
            \operatorname{logit}^{-1}\left(-2 + 0.05 \cdot (\mathbf{L}^1_t - L^1_{t-1})^2 + 0.1 \cdot L^2_{t-1} + 0.1 \cdot L^3_{t-1}  + \delta(A_{t-1})\right), &\\
        &\text{where }\delta(A_{t-1}) = 
        \begin{cases}
            -4, & \text{if } A_{t-1} \in \{\text{ari., halo., quet.}\}, \\
            -5, & \text{if } A_{t-1} = \text{risp.}, \\
            0, & \text{otherwise}.
        \end{cases} &
    \end{flalign*}
\end{itemize}

\subsubsection*{A.1.2.\enspace Treatment assignment}
The baseline treatment assignment ($A_0$) is given by \\
$$A_0 \mid \mathbf{V}^2, \mathbf{L}^1_0, \mathbf{L}^2_0, \mathbf{L}^3_0 \sim \text{Multinomial}\left(1; \mathbf{p}\right),$$ where $\mathbf{p}$ is a vector of weights for the initial treatment assignment ($A_0$), normalized to sum to one, given by:
\begin{align*}
\mathbf{p} = \begin{cases}
\frac{1}{4}, & \text{for Aripiprazole, if } \mathbf{V}^2 = \text{MDD} \text{ and } \mathbf{L}^3_0 > 0, \text{otherwise } \frac{1}{8} \\ 
\frac{1}{4}, & \text{for Haloperidol, if } \mathbf{V}^2 = \text{Schizophr.} \text{ and } \mathbf{L}^1_0 > 0, \text{otherwise } \frac{1}{8} \\
\frac{1}{4}, & \text{for Quetiapine, if } \mathbf{V}^2 = \text{Bipolar} \text{ and } \mathbf{L}^2_0 > 0, \text{otherwise } \frac{1}{8} \\
\frac{1}{4}, & \text{for Risperidone, if } \mathbf{V}^2 = \text{Bipolar} \text{ and } (\mathbf{L}^1_0 > 0 \text{ or } \mathbf{L}^2_0 > 0 \text{ or } \mathbf{L}^3_0 > 0), \text{otherwise } \frac{1}{8} \\
\frac{1}{8}, & \text{for Olanzapine and Ziprasidone}.
\end{cases}
\end{align*}

For treatment assignment at time $t>0$, let $\mathbf{P}_t$ be a $6\times 6$ transition probability matrix where $P_t[i,j]$ represents the probability of transitioning from treatment $i$ to treatment $j$:
\[
P_t[i,j] = \begin{cases}
0.95, & \text{if } i=j \text{ (staying on same treatment)} \\
0.01, & \text{if } i\neq j \text{ (switching treatments)}.
\end{cases}
\]
Then treatment is assigned according to:
\[
A_t \mid \bar{\mathcal{D}}_t \sim \text{Multinomial}(1; \mathbf{p}_t),
\]
where $\mathbf{p}_t$ is obtained by adding an adjustment vector $\mathbf{u}_t$ to the row of $\mathbf{P}_t$ corresponding to the previous treatment $A_{t-1}$. The elements of $\mathbf{u}_t$ are (0 when the stated condition does not hold):
\[
u_{t,j} = \begin{cases}
0.01, & \text{for } j=\text{ari.} \text{ if } \mathbf{L}^1_t > 0 \text{ or } \mathbf{L}^2_t > 0 \text{ or } \mathbf{L}^3_t > 0 \\
0.01, & \text{for } j=\text{halo.} \text{ if } \mathbf{L}^1_t > 0 \\
0.01, & \text{for } j=\text{quet.} \text{ if } \mathbf{L}^2_t > 0 \\
0.01, & \text{for } j=\text{risp.} \text{ if } \mathbf{L}^3_t > 0 \\
0, & \text{for } j \in \{\text{olan.}, \text{zipr.}\}.
\end{cases}
\]
The resulting vector is normalized to sum to one.

\subsubsection*{A.1.3.\enspace Censoring and outcome} 
Both the censoring event and the outcome are absorbing (implemented with \texttt{EFU = TRUE} in \texttt{simcausal}): once a subject is censored or has $Y_{t-1}=1$, no further draws are made and $Y_s = 1$ is carried forward after an event. At baseline, there is no censoring and $Y_0 \equiv 0$. Consequently, the term $Y_{t-1}$ below is 0 for every subject at risk.
For the \textbf{Monthly censored indicator} $C_t \mid \bar{\mathcal{D}}_t$ (in this subsection the Bernoulli draws are for the censoring event, i.e., for $1-C_t$ in the notation of Section~\ref{estimators}): \\
\begin{align*}
    C_t \mid \bar{\mathcal{D}}_t &\sim \text{Bernoulli}(p_t), \\
    p_t &=
    \begin{cases}
        1, & \text{if } \mathbf{V}^3 + \frac{t}{12} > 65, \\
        \operatorname{logit}^{-1}\left(\alpha + \beta(\mathbf{L}^1_t, \mathbf{L}^1_{t-1}, \mathbf{L}^2_{t-1}, \mathbf{L}^2_t, \mathbf{L}^3_{t-1}, \mathbf{L}^3_t, \bar{\mathbf{A}}_{t-1}, A_t)\right), & \text{otherwise},
    \end{cases}
\end{align*}
where \\
\begin{flalign*}
    &\alpha = -4, & \\
    &\beta(\mathbf{L}^1_t, \mathbf{L}^1_{t-1}, \mathbf{L}^2_{t-1}, \mathbf{L}^2_t, \mathbf{L}^3_{t-1}, \mathbf{L}^3_t, \bar{\mathbf{A}}_{t-1}, A_t) = & \\
    &0.1 \cdot (\mathbf{L}^1_t - \mathbf{L}^1_{t-1})^2 + 0.1 \cdot \mathbf{L}^2_{t-1} + 0.1 \cdot \mathbf{L}^2_t + 
    0.1 \cdot \mathbf{L}^3_{t-1} + 0.1 \cdot \mathbf{L}^3_t + \delta(\bar{\mathbf{A}}_{t-1}, A_t),&
\end{flalign*}
and for the \textbf{Diabetes} outcome $Y_t \mid \bar{\mathcal{D}}_t$: \\
\begin{align*}
    Y_t \mid \bar{\mathcal{D}}_t &\sim \text{Bernoulli}\left(\operatorname{logit}^{-1}\left(\alpha + Y_{t-1} + \beta(\mathbf{L}^1_t, \mathbf{L}^1_{t-1}, \mathbf{L}^2_{t-1}, \mathbf{L}^2_t, \mathbf{L}^3_{t-1}, \mathbf{L}^3_t, \bar{\mathbf{A}}_{t-1}, A_t)\right)\right).
\end{align*}
The treatment effect function $\delta(\bar{\mathbf{A}}_{t-1}, A_t)$ is defined as: \\
\begin{align*}
    \delta(\bar{\mathbf{A}}_{t-1}, A_t) &= 
    \begin{cases}
        -0.5, & \text{if } A_{t-1} \in \{\text{ari., halo., quet.}\}, \\
        -0.25, & \text{if } A_{t-1} = \text{risp.}, \\
        0, & \text{otherwise},
    \end{cases} \\
    &\quad + 
    \begin{cases}
        -1, & \text{if } A_t \in \{\text{ari., halo., quet.}\}, \\
        -2, & \text{if } A_t = \text{risp.}, \\
        0, & \text{otherwise}.
    \end{cases}
\end{align*}

\clearpage 
\bibliography{references}

\clearpage
\section*{Supporting Information (SI)}\label{SI}
\setcounter{table}{0}\renewcommand{\thetable}{S\arabic{table}}
\setcounter{figure}{0}\renewcommand{\thefigure}{S\arabic{figure}}

\begin{table}[htbp]
\centering
\caption{Hyperparameter settings for machine learning algorithms (T/C: treatment and censoring networks).}
\label{tab:hyperparams}
\begin{tabular}{lll}
\toprule
Algorithm & Parameter & Value \\
\midrule
\texttt{ranger} & Number of trees & 100 \\
\midrule
\multirow{2}{*}{\texttt{glmnet}} & $\alpha$ (lasso) & 1.0 \\
& $\alpha$ (elastic net) & 0.5 \\
& No. folds & 3 \\
& No. $\lambda$ values & 10 \\
\midrule
\multirow{18}{*}{LSTM} & \emph{Architecture} \\
& LSTM units per layer & 32 \\
& LSTM layers & 2 (T/C); 1 (outcome) \\
& Hidden activation & tanh \\
& Recurrent activation & sigmoid \\
& Dense activation & ReLU \\
& Self-attention & Single-head, causal (T/C); none (outcome) \\
& Pre-output dense layer units & 16 (with input skip) \\
\multirow{19}{*}{ } \\
\multirow{18}{*}{ } & \emph{Training} \\
& Batch size & 128 \\
& Maximum epochs & 30 (T/C); 20 / 6 (outcome) \\
& Early stopping patience & 3 (T/C); 3 / 2 (outcome) \\
& Dropout rate & 0.1 \\
& Weight decay (\texttt{AdamW}) & $1 \times 10^{-4}$ \\
& Epsilon & $1 \times 10^{-7}$ \\
& Gradient clipping (norm) & 1.0 \\
& Window size (outcome) & 12 \\
\multirow{19}{*}{ } \\
\multirow{18}{*}{ } & \emph{Data splits} \\
& Training split & 90\% (subject-level) \\
& Validation split & 10\% (subject-level) \\
\multirow{19}{*}{ } \\
\multirow{18}{*}{ } & \emph{Learning rate} \\
& Constant learning rate & 0.001 (T/C); 0.003 (outcome) \\
\bottomrule
\end{tabular}
\end{table}

\subsection*{Performance metrics used in simulations}\label{performance-metrics-supporting}
We consider four metrics to evaluate each estimator's ability to recover the true parameter value $\psi_t$. The first metric focuses on bias.

\begin{definition}
{\em \textbf{Mean absolute bias}. The mean absolute difference between the true parameter value and the estimated parameter value at time $t$ under treatment rule $\bar{\mathbf{d}}_t$, averaged over $R$ simulation runs:}
\begin{equation*}
\text{Absolute bias} = \frac{1}{R} \sum_{r=1}^{R} \left| \hat{\psi}_{t}^{(r)} - \psi_{t} \right|,
\end{equation*}
where $\psi_{t}$ is the true parameter value, computed once from a large counterfactual sample.
\end{definition}

The second metric assesses the performance of the influence curve-based confidence intervals.
\begin{definition}
{\em \textbf{Coverage probability}. The proportion of estimated 95\% confidence intervals in the $R$ simulation runs that contain the true parameter value:}
\begin{equation*}
\text{Coverage probability} \;=\; \frac{1}{R} \sum_{r=1}^{R} \mathbbm{1}\!\Big\{\,\psi_{t} \in \Big[\;\hat{\psi}_{t}^{(r)} \pm 1.96\,\sqrt{\tfrac{\widehat{\operatorname{Var}}(\widehat{IC}(\psi_t))}{n}}\;\Big]\Big\}~.
\end{equation*}
\end{definition}

The third metric quantifies the precision of the estimates.
\begin{definition}
{\em \textbf{Confidence interval width}. The average width of the estimated 95\% confidence intervals over the $R$ simulation runs:}
\begin{equation*}
\text{Confidence interval width} = \frac{1}{R}\sum_{r=1}^{R} 2 \times 1.96 \sqrt{\frac{\widehat{\operatorname{Var}}(\widehat{IC}(\psi_t))}{n}}.
\end{equation*}
\end{definition}

The fourth metric evaluates the data support for a given intervention rule, serving as a diagnostic tool for positivity violations.

\begin{definition}
{\em \textbf{Proportion with low cumulative treatment probability}. The estimated cumulative probability of continuing to receive treatment according to the assigned treatment rule through time \(t\), \(\prod_{s=0}^{t} \hat{P}(A_s = d_s \mid \bar{\mathbf{L}}_s = \bar{\mathbf{l}}_s, \bar{\mathbf{A}}_{s-1} = \bar{\mathbf{d}}_{s-1})\), reported as the proportion of the $n_t^{(r)}$ uncensored rule-followers at time $t$ (set $\mathcal{F}_t^{(r)}$) with cumulative probabilities smaller than 0.025, averaged over \(R\) simulation runs, indicating potential positivity violations. For the stochastic rule, all uncensored subjects are rule-followers and the product is replaced by the cumulative ratio $\prod_{s=1}^{t} \hat{P}(A_s \mid \cdot)/g^*_s(A_s \mid A_{s-1})$:}
\begin{equation*}
\Pr\text{(low support)} \;=\; \frac{1}{R}\sum_{r=1}^{R}\frac{1}{n_t^{(r)}}\sum_{i \in \mathcal{F}_t^{(r)}}\mathbbm{1}\Big\{\prod_{s=0}^{t}\hat{\Pr}\big(A_s=d_s\mid\bar{\mathbf{L}}_s,\bar{\mathbf{A}}_{s-1}=\bar{\mathbf{d}}_{s-1}\big)_i^{(r)}<0.025\Big\}~.
\end{equation*}
\end{definition}

\begin{figure}[htbp]
    \centering
    \begin{subfigure}[b]{0.49\textwidth}
        \includegraphics[width=\textwidth]{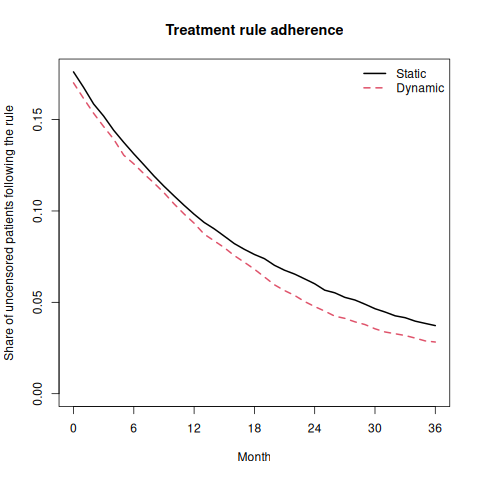}
        \caption{Treatment rule adherence over time among uncensored patients (static and dynamic rules).}
        \label{fig:adherence}
    \end{subfigure}
    \hfill
    \begin{subfigure}[b]{0.49\textwidth}
        \includegraphics[width=\textwidth]{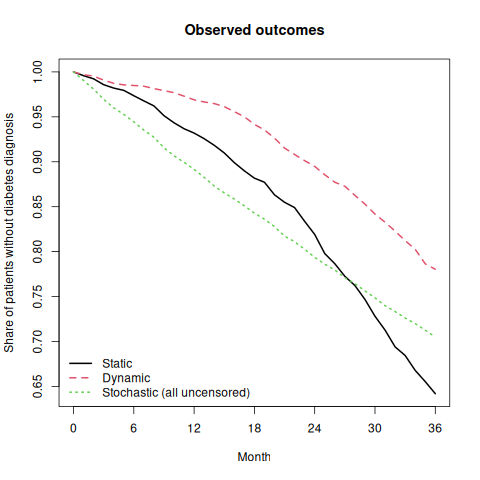}
        \caption{Observed survival curves (proportion without diabetes diagnosis) with censoring.}
        \label{fig:survival_observed}
    \end{subfigure}
    \hfill
    \begin{subfigure}[b]{0.49\textwidth}
        \includegraphics[width=\textwidth]{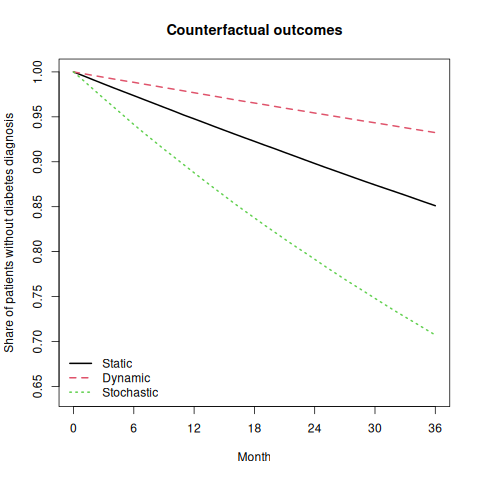}
        \caption{True counterfactual survival curves without censoring and under perfect treatment adherence.}
        \label{fig:survival_truth}
    \end{subfigure}
    \caption{Visualization of treatment adherence patterns and their impact on observed versus counterfactual survival outcomes in a single simulation run with $n=10000$ patients. The stochastic rule assigns a positive probability to every treatment path, so all uncensored patients are compatible with it; its curve in panel (b) (dotted) is therefore the marginal survival curve among uncensored patients in the observed data.}
    \label{fig:combined_simulation}
\end{figure}

\begin{table}[ht]
\centering
\caption{Performance metrics by estimator and treatment rule, aggregated across time points. Values are over 100 simulation runs ($n=10{,}000$) and target time points $t=12, 24, 36$. Coverage probability: proportion of 95\% CIs containing the true value (Monte Carlo SE); CI widths: mean (median); absolute bias: mean $\pm$ SD over all replicate-by-cell values. G-computation intervals are approximate (influence curve evaluated at the untargeted fits).}
\label{tab:results-estimator-rule}
\resizebox{0.8\textwidth}{!}{%
\begin{tabular}{ll|c|c|c}
\hline
\multicolumn{2}{c|}{\textbf{Group}} & \textbf{Coverage prob.} & \textbf{CI Widths} & \textbf{Abs. Bias} \\
\hline
G-Comp-SL & Static & $0.910$ (0.017) & $0.0543$ (0.0512) & $0.0190 \pm 0.0088$ \\ 
G-Comp-SL & Dynamic & $0.013$ (0.007) & $0.0337$ (0.0305) & $0.0380 \pm 0.0178$ \\ 
G-Comp-SL & Stochastic & $0.840$ (0.021) & $0.0175$ (0.0179) & $0.0054 \pm 0.0042$ \\ 
IPTW-SL (bin.) & Static & $0.873$ (0.019) & $0.0442$ (0.0422) & $0.0116 \pm 0.0107$ \\ 
IPTW-SL (bin.) & Dynamic & $0.933$ (0.014) & $0.0342$ (0.0320) & $0.0070 \pm 0.0080$ \\ 
IPTW-SL (bin.) & Stochastic & $0.943$ (0.013) & $0.0186$ (0.0190) & $0.0038 \pm 0.0031$ \\ 
IPTW-SL (multi.) & Static & $0.813$ (0.022) & $0.0442$ (0.0420) & $0.0131 \pm 0.0121$ \\ 
IPTW-SL (multi.) & Dynamic & $0.933$ (0.014) & $0.0344$ (0.0321) & $0.0071 \pm 0.0081$ \\ 
IPTW-SL (multi.) & Stochastic & $0.673$ (0.027) & $0.0180$ (0.0183) & $0.0075 \pm 0.0055$ \\ 
LTMLE-SL (bin.) & Static & $0.963$ (0.011) & $0.0502$ (0.0479) & $0.0088 \pm 0.0078$ \\ 
LTMLE-SL (bin.) & Dynamic & $0.917$ (0.016) & $0.0317$ (0.0301) & $0.0065 \pm 0.0072$ \\ 
LTMLE-SL (bin.) & Stochastic & $0.923$ (0.015) & $0.0169$ (0.0173) & $0.0036 \pm 0.0030$ \\ 
LTMLE-SL (multi.) & Static & $0.973$ (0.009) & $0.0540$ (0.0510) & $0.0091 \pm 0.0080$ \\ 
LTMLE-SL (multi.) & Dynamic & $0.920$ (0.016) & $0.0317$ (0.0301) & $0.0065 \pm 0.0072$ \\ 
LTMLE-SL (multi.) & Stochastic & $0.927$ (0.015) & $0.0175$ (0.0179) & $0.0037 \pm 0.0029$ \\ 
\hline
G-Comp-RNN & Static & $1.000$ (0.000) & $0.0436$ (0.0425) & $0.0053 \pm 0.0040$ \\ 
G-Comp-RNN & Dynamic & $0.967$ (0.010) & $0.0288$ (0.0272) & $0.0056 \pm 0.0045$ \\ 
G-Comp-RNN & Stochastic & $0.820$ (0.022) & $0.0171$ (0.0175) & $0.0053 \pm 0.0040$ \\ 
IPTW-RNN (bin.) & Static & $0.933$ (0.014) & $0.0479$ (0.0454) & $0.0110 \pm 0.0110$ \\ 
IPTW-RNN (bin.) & Dynamic & $0.920$ (0.016) & $0.0371$ (0.0347) & $0.0093 \pm 0.0106$ \\ 
IPTW-RNN (bin.) & Stochastic & $0.843$ (0.021) & $0.0181$ (0.0183) & $0.0059 \pm 0.0130$ \\ 
IPTW-RNN (multi.) & Static & $0.947$ (0.013) & $0.0471$ (0.0451) & $0.0104 \pm 0.0094$ \\ 
IPTW-RNN (multi.) & Dynamic & $0.913$ (0.016) & $0.0372$ (0.0343) & $0.0097 \pm 0.0110$ \\ 
IPTW-RNN (multi.) & Stochastic & $0.843$ (0.021) & $0.0178$ (0.0183) & $0.0051 \pm 0.0043$ \\ 
LTMLE-RNN (bin.) & Static & $0.940$ (0.014) & $0.1012$ (0.0455) & $0.0093 \pm 0.0081$ \\ 
LTMLE-RNN (bin.) & Dynamic & $0.893$ (0.018) & $0.0298$ (0.0285) & $0.0067 \pm 0.0074$ \\ 
LTMLE-RNN (bin.) & Stochastic & $0.883$ (0.019) & $0.0187$ (0.0175) & $0.0041 \pm 0.0034$ \\ 
LTMLE-RNN (multi.) & Static & $0.947$ (0.013) & $0.0436$ (0.0424) & $0.0087 \pm 0.0075$ \\ 
LTMLE-RNN (multi.) & Dynamic & $0.897$ (0.018) & $0.0283$ (0.0272) & $0.0066 \pm 0.0073$ \\ 
LTMLE-RNN (multi.) & Stochastic & $0.897$ (0.018) & $0.0171$ (0.0176) & $0.0040 \pm 0.0034$ \\ 
\hline
\end{tabular}
}
\end{table}

\begin{table}[h]
\centering
\caption{Proportion of uncensored rule-followers whose estimated cumulative probability of following the rule is smaller than 0.025, for static, dynamic, and stochastic treatment rules under the SL and RNN implementations (mean over 100 simulation runs; multinomial treatment model). For the stochastic rule, the cumulative ratio of the estimated to the rule's treatment probabilities is used.}
\label{tab:positivity}
\begin{tabular}{l|ccc}
\toprule
\multicolumn{4}{l}{\textbf{Static Rule}} \\ 
\textbf{Implementation} & $t=12$ & $t=24$ & $t=36$ \\ 
\midrule
SL & 0.000 & 0.008 & 0.925 \\ 
RNN & 0.000 & 0.000 & 0.382 \\ 
\midrule
\multicolumn{4}{l}{\textbf{Dynamic Rule}} \\ 
\textbf{Implementation} & $t=12$ & $t=24$ & $t=36$ \\ 
\midrule
SL & 0.000 & 0.000 & 0.954 \\ 
RNN & 0.000 & 0.000 & 0.269 \\ 
\midrule
\multicolumn{4}{l}{\textbf{Stochastic Rule}} \\ 
\textbf{Implementation} & $t=12$ & $t=24$ & $t=36$ \\ 
\midrule
SL & 0.000 & 0.000 & 0.000 \\ 
RNN & 0.000 & 0.000 & 0.000 \\ 
\bottomrule
\end{tabular}
\end{table}

\begin{table}[ht]
\centering
\caption{Computation time per simulation run (minutes; median and interquartile range over 100 runs with $n=10{,}000$ and target time points $t=12, 24, 36$). Treatment and censoring models: multinomial and binomial treatment models and censoring models at all time points; outcome regressions: the backward recursions with targeting for all target times and rules. The LTMLE, IPTW and g-computation estimates of a run share these fits. CPU time includes the worker processes. Each replicate used 3 worker processes on a laptop with a 4-core Intel Core i7-8550U CPU (no GPU).}
\label{tab:run-times}
\resizebox{\textwidth}{!}{%
\begin{tabular}{l|cccc|c}
\hline
 & \multicolumn{3}{c}{\textbf{Wall clock}} & \textbf{CPU time} & \textbf{Total} \\
\textbf{Implementation} & \textbf{Treatment, censoring} & \textbf{Outcome regressions} & \textbf{Run} & \textbf{Run} & \textbf{(hours)} \\
\hline
SL & 38.7 (37.6--40.3) & 17.6 (17.2--18.1) & 56.5 (55.2--58.3) & 161.9 (158.7--167.6) & 95.3 \\ 
RNN & 4.5 (4.1--4.9) & 20.6 (20.3--21.3) & 25.2 (24.6--26.3) & 63.8 (62.8--65.9) & 42.8 \\ 
\hline
\end{tabular}
}
\end{table}

\begin{figure}[htbp]
    \centering
    \begin{subfigure}[b]{0.49\textwidth}
        \includegraphics[width=\textwidth]{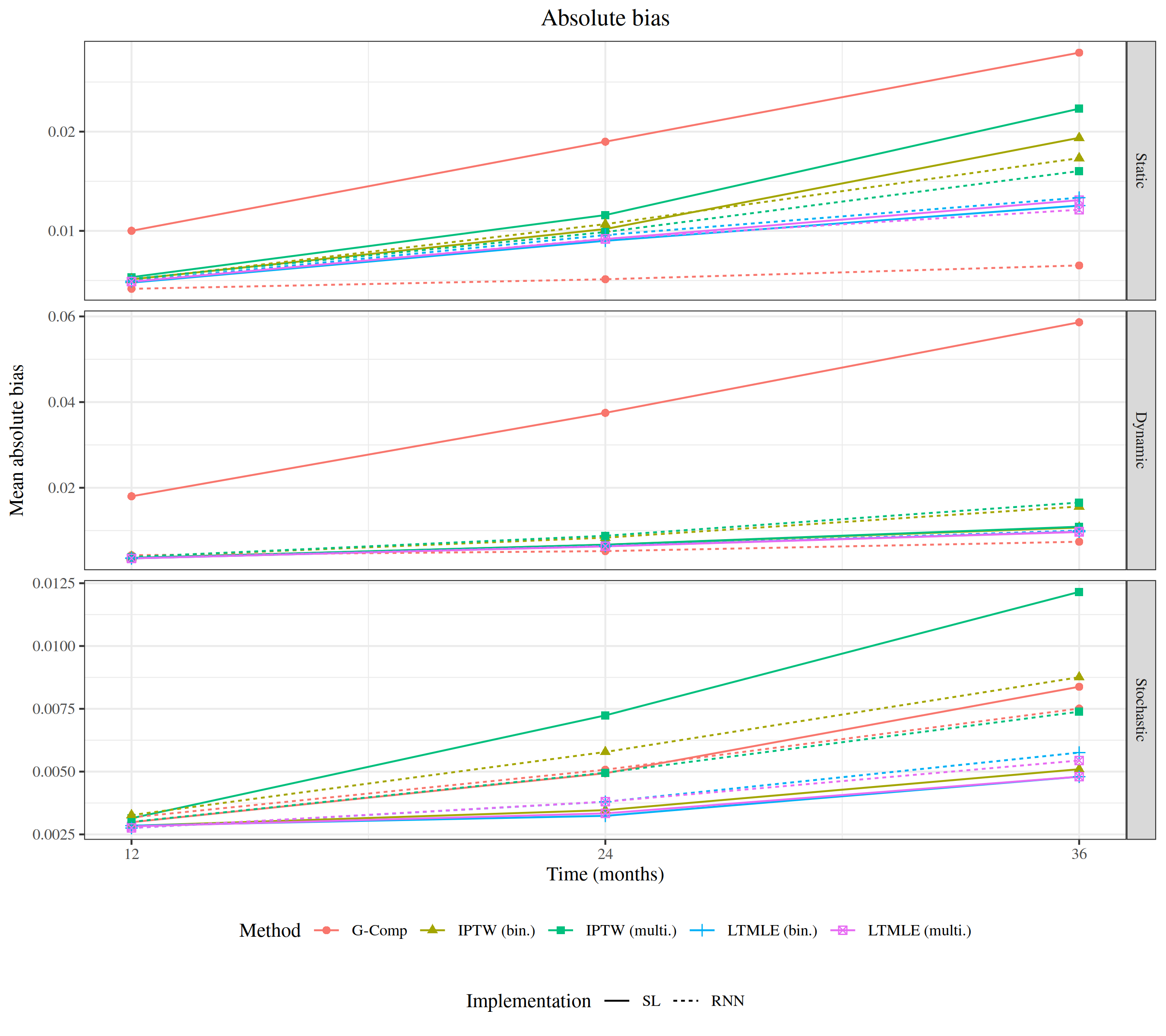}
        \caption{Absolute bias}
        \label{fig:bias}
    \end{subfigure}
    \hfill
    \begin{subfigure}[b]{0.49\textwidth}
        \includegraphics[width=\textwidth]{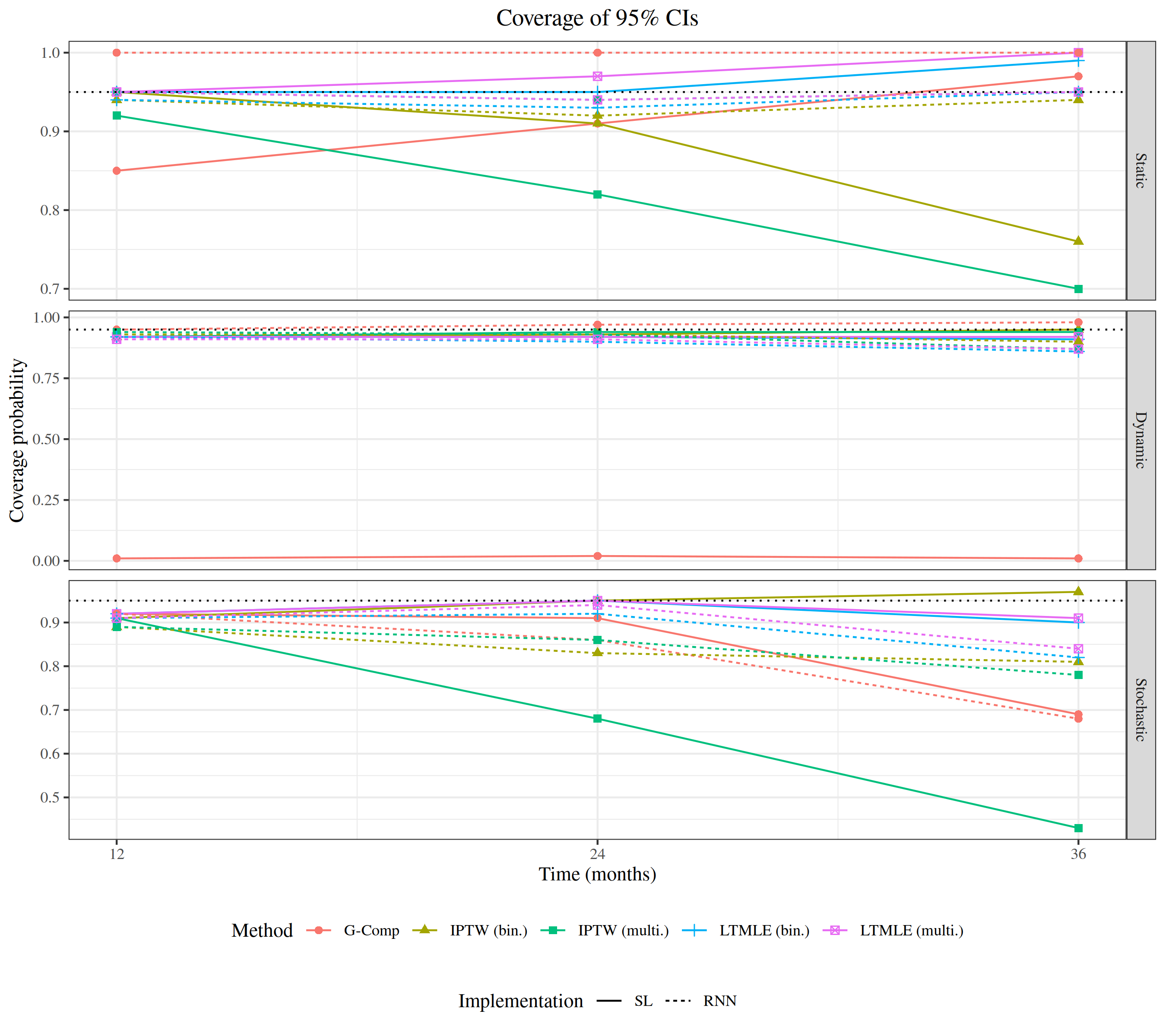}
        \caption{Coverage probability}
        \label{fig:coverage}
    \end{subfigure}
    \hfill
    \begin{subfigure}[b]{0.49\textwidth}
        \includegraphics[width=\textwidth]{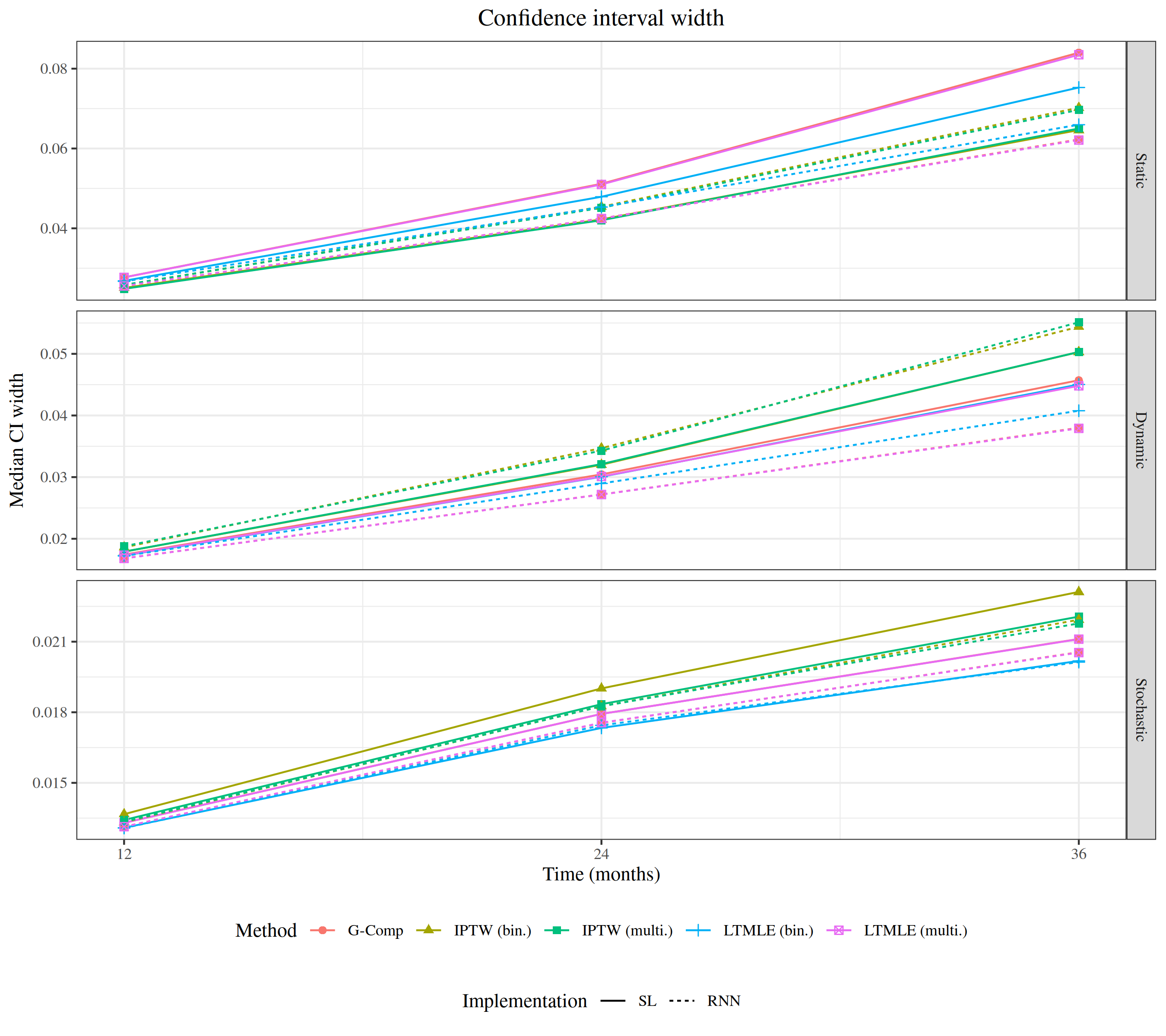}
        \caption{Confidence interval width}
        \label{fig:ci_width}
    \end{subfigure}
    \caption{Comparative analysis of RNN-- and SL-based estimators over time based on 100 simulation runs with $n=10000$ patients, at target time points $t=12, 24, 36$. Panel (a) shows the mean absolute bias between true and estimated counterfactual outcomes for each treatment rule. Panel (b) illustrates the coverage probability (proportion of runs) of 95\% confidence intervals (dotted line shows nominal level). Panel (c) shows the median confidence interval widths. Results are shown for LTMLE (multi. and bin.), IPTW (multi. and bin.), and G-Comp estimators with SL (solid lines) and RNNs (dashed lines) under static, dynamic, and stochastic treatment rules; g-computation intervals are approximate.}
    \label{fig:sim_results_long}
\end{figure}

\end{document}